\documentclass[manuscript]{aastex}

\slugcomment{To appear in ApJ}
\shorttitle{A.-P. cosmological test}
\shortauthors{L\'opez-Corredoira}

\usepackage{epsfig}
\usepackage{graphicx,lscape}
\usepackage{amsmath,amsfonts,amssymb}
\newcommand{\be}{\begin{equation}}
\newcommand{\ee}{\end{equation}}
\newcommand{\bea}{\begin{eqnarray}}
\newcommand{\eea}{\end{eqnarray}}

\begin{document}

\title{Alcock-Paczy\'nski cosmological test}

\author{M. L\'opez-Corredoira\altaffilmark{1,2}}
\affil{$^1$ Instituto de Astrof\'\i sica de Canarias,
E-38200 La Laguna, Tenerife, Spain\\
$^2$ Departamento de Astrof\'\i sica, Universidad de La Laguna,
E-38206 La Laguna, Tenerife, Spain}
\email{martinlc@iac.es}

\begin{abstract}
In order to test the expansion of the universe and its geometry, we 
carry out an Alcock \& Paczy\'nski cosmological test, that
is, an evaluation of the ratio of observed angular size to radial/redshift size.
The main advantage of this test is that it does not depend on the evolution of the galaxies, but only on the geometry of the universe. However, the redshift distortions produced by the peculiar velocities of the gravitational infall do also have an influence, which should be separated from the cosmological effect.
We derive the anisotropic correlation function of sources in three surveys within the Sloan Digital Sky Survey (SDSS): galaxies from SDSS-III/Baryon Oscillation Spectroscopy Survey--Data Release 10 (BOSS-DR10), and QSOs from SDSS-II and SDSS-III/BOSS-DR10. From these, we are able to disentangle the dynamic and geometric distortions and thus derive the ratio of observed angular size to radial/redshift size at different redshifts. We also add some other values available in the literature. Then, we use the data to evaluate which cosmological
model fits them. We used six different models: concordance $\Lambda $CDM, Einstein-de Sitter, open--Friedman Cosmology without dark energy, flat quasi-steady state cosmology, a static universe with a linear Hubble law, and a static universe with tired--light redshift. Only two of the six models above fit the data of the Alcock \& Paczy\'nski test: concordance $\Lambda $CDM and static universe with tired--light redshift; whereas the rest of them are excluded at a $>95$\% confidence level. If we assume that $\Lambda $CDM is the correct one, the best fit with a free $\Omega _m$ is produced for $\Omega _m=0.24^{+0.10}_{-0.07}$.
\end{abstract}


\section{Introduction}

There are several ways to test the geometry of a cosmological model
and its expansion, i.e., that redshifts of the
galaxies are cosmological and not due to an alternative
mechanism (Narlikar 1989; L\'opez-Corredoira 2003, Section 2.1; L\'opez-Corredoira 2006). However, almost all of the cosmological
tests are entangled with the evolution of galaxies and/or other effects. Hubble diagrams (LaViolette 1986; Schade et al. 1997; Marosi 2013), Tolman surface--brightness tests (Lubin \& Sandage 2001; Andrews 2006; Lerner 2006), angular size tests (Kapahi 1987; Kellerman 1993; L\'opez-Corredoira 2010) are all affected by the evolution of galaxies at high redshifts. So far, there are no standard rods in physical objects. The 
ultra-compact radio sources which were claimed in the past to be free of evolutionary effects are now thought to present some evolution as well (L\'opez-Corredoira 2010; Pashchenko \& Vitrishchak 2011). L\'opez-Corredoira (2010) thinks that the huge size evolution necessary to fit an angular size test with an expanding universe is not understood, but it still depends on our understanding of the galaxies rather than on pure cosmological approaches. Baryonic acoustic oscillations (BAOs) in the cosmic microwave background radiation (CMBR) or in the large scale structure are usually thought to be standard rods to measure the cosmological expansion (e.g., Rassat \& Refregier 2012); however their features in the power spectrum are not a direct observable quantity free of assumptions (one must assume at least that BAO peaks represent a faint imprint of the sound waves in the clustering of galaxies and matter today excited by the initial inflationary perturbations, i.e. the assumption that the scenario of standard cosmology is correct) and these peaks could be generated with other cosmological assumptions different from the standard interpretation of acoustic peaks (L\'opez-Corredoira \& Gabrielli 2013).

Time dilation tests in Type Ia supernovae (SNIa) look like one of the most successful tests in favor of the expansion of the universe (Goldhaber et al. 2001; Blondin et al. 2008), but they do not serve to test the geometry of the universe, and there are still some problems in the interpretation. The fact that SNe Ia light curves are narrower when redder (Nobili \& Goobar 2008) is an inconvenience for a clean test free of selection effects. Other selection effects and the possible compatibility of the results with a wider range of cosmological models, including static ones, were also pointed out by L\'opez-Corredoira (2003, Section 2 and references therein), Brynjolfsson (2004b); Leaning (2006); Crawford (2009b, 2011); Holushko (2012), and 
LaViolette (2012, Section 7.8).
Moreover, neither gamma-ray bursts (Crawford 2009a) nor QSOs (Hawkins 2010) present time dilation, which is puzzling.

There are, of course, the CMBR anisotropies as a way to test cosmological models; they are, perhaps, the most important support for the standard model. However, one should find an independent confirmation, because it is possible to generate/modify CMBR anisotropies by mechanisms different than the standard cosmology (Narlikar et al. 2007; Angus \& Diaferio 2011; L\'opez-Corredoira \& Gabrielli 2013; L\'opez-Corredoira 2013) and/or contamination (L\'opez-Corredoira 2007). 

The microwave background temperature measured from
rotational excitation of some molecules as a function of redshift (Molaro et al. 2002; Noterdaeme et al. 2011) is another possible test and was quite successful in proving the expansion: results from 
Noterdaeme et al. (2011) with the exact expected dependence of
$T=T_0(1+z)$ are impressive. Nonetheless, there are other results which disagree 
with that dependence (Krelowski et al. 2012; Sato et al. 2013). It might be due to
a dependence on collisional excitation (Molaro et al.
2002) or bias due to unresolved structure (Sato et al. 2013).

The temperature of the intergalactic medium as a function of redshift
might also be useful to constrain more directly the geometry of the universe.
At present, it is typically inferred to be 
20,000 K; there is no evidence of evolution with redshift (Zaldarriaga et al. 2001), which is puzzling in an expanding universe.

Here, we propose an Alcock \& Paczy\'nski (1979) cosmological test:
an evaluation of the ratio of observed angular size to radial/redshift size
(see detailed explanation in Section \ref{.method}).
The main advantage of this test is that it does not depend on the evolution of the galaxies, but only on the geometry of the universe. However, the redshift distortions (Kaiser 1987; Hamilton 1998) do also have
an influence. Although there already have been many attempts to carry out this test (see references in Section \ref{.2pcf}), here we adapt the method (Sections \ref{.method} and \ref{.2pcf_calc}) and focus on the application as a test for any cosmological model, whereas other authors
assume the standard cosmology and just fit some of their parameters.
We apply it to three spectroscopic surveys with a total of $\sim 10^6$ galaxies up to redshift 0.8 and $\sim 10^5$ QSOs at higher redshifts (Sections \ref{.application} and \ref{.fitcosmo}). 
As will be discussed in Section \ref{.conclusions}, this test will give us information about which cosmological model fits better. Hence, this is a cosmological test free of entanglement with evolution, for checking both the geometry and the expansion of the universe. Certainly, this exercise is not going to solve the question once and for all, but the method is a possible way to do it with the increasing possibility of future huge spectroscopic surveys.

\section{Method of the test}
\label{.method}

The key point of the test is based on the fact that the two-point correlation function in a
distribution of galaxies in the real space must have spherical symmetry, that is, 
it only depends on the distance between the sources and not
on any of the two angles which define the relative position between the two sources.
This means that the fall of the two-point correlation function as a function of the distance 
must be the same either along the line of sight ($r_\parallel $) or perpendicular to it ($r_\perp $). The distance of two sources along the line of sight is given by the difference of redshift, $\Delta z$, and depends on the cosmological model. The distance of two sources in the direction perpendicular to the line of sight
is given by the angular separation, $\Delta \theta $, and also depends on the cosmological model; this is precisely what gives meaning to the angular-size tests.

\subsection{Dependence on the Cosmological Model}
 
\begin{figure}
\vspace{1cm}
{\par\centering \resizebox*{8cm}{8cm}{\includegraphics{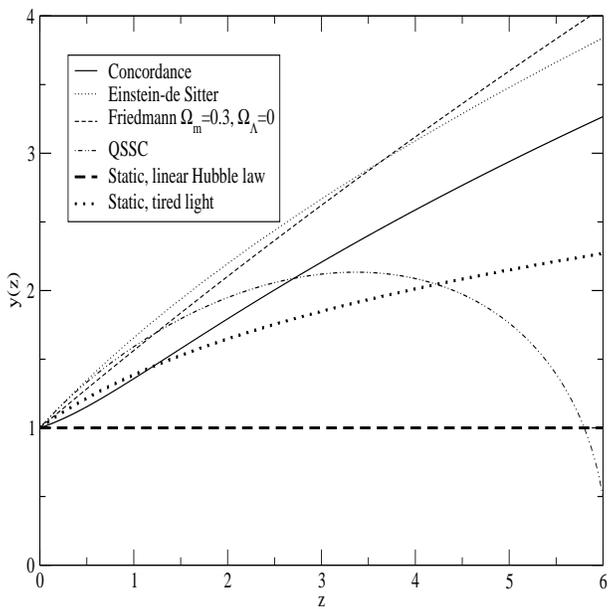}}
\par\centering}
\caption{Values of the function $y(z)$ for the six cosmological models given in 
Appendix \S \ref{.cosmomodels} with the parameters specified in that section.}
\label{Fig:models}
\end{figure}

The comoving distance between two sources at redshift $z$ separated by a relatively small
$\Delta z$ and $\Delta \theta $ (radians) in the redshift-space is (see Appendix \S \ref{.cosmomodels})
\begin{equation}
s(\Delta z, z\Delta \theta )\approx \frac{c}{H_0}x(z)\sqrt{(\Delta z)^2\ +\ y^2(z)\ (z\Delta \theta )^2}
\label{comdist}
,\end{equation}
where $c$ is the speed of light, $H_0$ is the Hubble constant, and $x(z)$ and $y(z)$ are functions which
depend on the cosmological model. For the standard concordance model (with the equation of
state for the dark energy $\omega _\Lambda =-1$):
\begin{equation}
y(z)=\frac{1}{z}\int _0^zdx \sqrt{\frac{\Omega _m(1+z)^3
+\Omega _\Lambda }{\Omega _m(1+x)^3
+\Omega _\Lambda }}
\label{yz_std}
,\end{equation}
\begin{equation}
x(z)=\frac{1}{\sqrt{\Omega _m(1+z)^3+\Omega _\Lambda }}
.\end{equation}
See Appendix \ref{.cosmomodels} for their dependence in other cosmological models; they are plotted
in Figure \ref{Fig:models}.

All models have in common that $y(0)=1$, but there are already some differences for low z there are some differences; in Appendix \ref{.cosmomodels}, we give the limits of $z\rightarrow 0$ for $y(z)$. For high $z$, the differences are even higher. 
The possibility of deriving $y(z)$ free of selection effects, evolution, redshift distortion, etc., 
would tell us which theoretical model is correct. 

Furthermore, note that even assuming the standard model as correct, the
derivation of $y(z)$ would allow us to determine $\Omega _m$ and $\Omega _\Lambda $.
For low $z$ in the concordance model (Equation (\ref{lowz_1})): 
$\lim _{z\rightarrow 0}y(z)\approx 1+\frac{3}{4}\Omega _mz$, so a high--precision
measurement of $y(z)$ in the local universe could give us a direct measurement of $\Omega _m$
(and consequently $\Omega _\Lambda $, assuming $\Omega _m+\Omega _\Lambda =1$).
This measurement of $y(z)$ is possible, as we will see next.

\subsection{Two-point Correlation Function}
\label{.2pcf}

As said, the two-point correlation function of
a distribution of galaxies ($\xi $) should only depend on the comoving distance, $r$, in the real space 
for a given redshift, i.e., $\xi=\xi(r;z)$. For the redshift-space ($s$), there is no 
spherical symmetry because we also have the redshift-space distortions in the field $\xi $ produced by the
peculiar velocities of the gravitational infall. In the linear regime of fluctuations, assuming $\xi(r;z)=A(z)r^{-\gamma (z)}$ (this is an acceptable approximation for scales $\lesssim 100$ $h^{-1}$Mpc, e.g., Ross et al. 2007; Sylos Labini et al. 2009; see also Figure \ref{Fig:xi0}) with distortion parameter $\beta (z)$,
\begin{equation}
\xi (\Delta z, z\Delta \theta ;z)=\xi(r;z)f[\mu ,\beta (z), \gamma (z)]
,\end{equation} 
where $\mu $ is the cosine of the angle between $s=\sqrt{s_\perp^2+s_\parallel ^2}$ and $s_\parallel =\frac{c}{H_0}x(z)\Delta z$; $s_\perp=\frac{c}{H_0}x(z)y(z)z\Delta \theta $;
and the function $f$ is (Matsubara \& Suto 1996):
\begin{equation}
f(\mu ,\beta, \gamma )=1+\frac{2(1-\gamma \mu ^2)}{3-\gamma}\beta +
\frac{3-6\gamma \mu ^2+\gamma (2+\gamma )\mu ^4}{(3-\gamma )(5-\gamma )}
\beta ^2
.\end{equation}
Hence, in polar coordinates $R=\sqrt{(\Delta z)^2+(z\Delta \theta )^2}$, $\alpha
=\cos^{-1}\left(\frac{\Delta z}{R}\right)$:
\begin{equation}
\label{xi}
\xi (R, \alpha ;z)=K(z)R^{-\gamma (z)}(\cos ^2\alpha
+y^2(z)\sin ^2\alpha )^{-\gamma (z)/2}
\end{equation}\[
\times f[\mu ,\beta (z), \gamma (z)]
,\]\[
\mu =\frac{1}{\sqrt{1+y^2(z)\tan ^2\alpha}}
,\]
where $K(z)$ stands for certain amplitude.

Note that we do not use $\xi (s_\perp, s_\parallel )$, like most of the references
that talk about the anisotropic two-point correlation function, because we are not
assuming any cosmology a priori. The advantage of the expression in Equation (\ref{xi}) is
that it relates an observable function totally free of any modeling to the
cosmology implicit in the values of $y(z)$, and the redshift-space distortions
implicit in $\beta (z)$. If we wanted to calculate $\xi (s_\perp, s_\parallel )$, we
would have to obtain $s_\perp $ and $s_\parallel $ using a given cosmological model and
change/rescale this calculation every time we wanted to test another model.

\begin{figure}
\vspace{1cm}
{\par\centering \resizebox*{7cm}{18.8cm}{\includegraphics{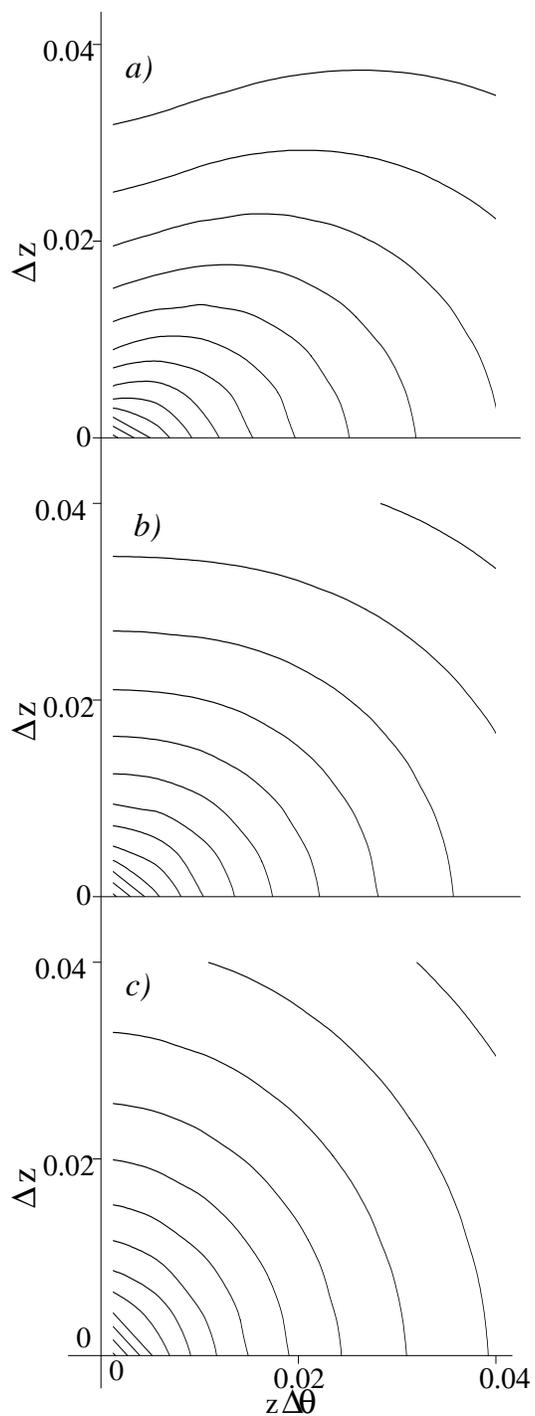}}
\par\centering}
\caption{Function $\log_{10}\xi (\Delta z, z\Delta \theta )$ for a model with $\gamma =2$ and $y=1.1$
for three different values of the redshift distortion: (a) $\beta =0.3$, (b) $\beta =0.15$, (c) $\beta =0$. 
The amplitudes are arbitrary; the contour step is 0.2 dex. Pixels are binned at $0.004\times 0.004$.}
\label{Fig:elipses}
\end{figure}

In Figure \ref{Fig:elipses}, the effect of the redshift distortion is illustrated.
If we had no redshift distortions ($\beta =0 \Rightarrow f=1$), $\xi (\Delta z, z\Delta \theta ;z)$ would
not depend on $\mu $ and, for a fixed $z$, when we plotted $\xi $ in the plane of $z\Delta \theta $,
$\Delta z$, the isocontours would be ellipses with axial ratios $y(z)$ of the 
second/first axes (Figure \ref{Fig:elipses}c). For the general case, we must consider redshift distortions.
The parameter $\beta (z)$ takes into account, as said, the large-scale effects of linear $z$-space distortions (Hamilton 1998).
Usually, in a standard scenario, it is related to cosmological information as $\beta (z)\approx 
\frac{\Omega _m(z)^{0.6}}{b(z)}$ (Kaiser 1987), 
where $b(z)$ is the bias parameter, but here, in general and for any cosmological
scenario, $\beta $ will be simply a parameter from which we will not extract any further information.
We must be aware that Kaiser's formalism was developed for the standard cosmology including expansion; 
however, this remains valid for any model, provided that the meaning of $\beta $ is changed. In general, it is valid when the
gravitational infall is given such that $\nabla \vec{v}=-\beta \delta $, where $\vec{v}$ is the velocity field
and $\delta $ the matter overdensity (in the linear regime); this yields even for a static universe. 
The continuity and Euler equation remain the same thing in all models, but the Poisson
equation changes with the gravitational model as well as the evolution of the linear growth of overdensities, thus giving rise to different equivalences of $\beta $ with the cosmological parameters.

The small-scale random motion of the galaxies should also be included in an analysis with small bin size:  we should convolve Equation (\ref{xi}) with a Gaussian distribution of $\Delta z$ (e.g., Ross et al. 2007, 
Section 4.2) to get the correct $\xi (\Delta z, z\Delta \theta ;z)$; however, we will neglect this correction since the dispersion of velocities is $\langle \omega_ z\rangle ^{1/2}\approx $ 300 km s$^{-1}$ (Ross et al. 2007), equivalent to $\Delta z\approx 0.001$, which is much smaller than our bin sizes.
The low values of $\Delta \theta $ in which
the features known as the ``fingers of God'' (Jackson 1972) are observed are not
explored here.

Obtaining $y(z)$ for different values of $z$ and comparing it with theoretical predictions of different cosmological models (Figure \ref{Fig:models}), perhaps allowing for some variation of $\Omega _m$ instead of a fixed value, is theoretically a good way to constrain cosmology.
Nonetheless, a difficulty arises from the quasi-degeneracy between the variations
of $y(z)$ and $\beta (z)$, i.e. between geometric and dynamic distortions, respectively. 
The range of possible cosmological models compatible with the data
for some free $\beta $ is too large (e.g., Ross et al. 2007), thus making it difficult to disentangle geometric and dynamic distortions.
There are some methods to obtain $\beta (z)$ from $\xi $ (e.g., Hamilton 1992, 1998; Tocchini-Valentini et al. 2012) or from $\langle |\mu |\rangle $ in a
sample of galaxies (Patiri et al. 2012), but they depend on $y(z)$. 
Nonetheless, there are ways to attack the problem. Some approaches to our problem try to fit $\beta $ for the different cosmological models [that is, fixing $y(z)$] and adopt as valid cosmological the one which gives
better fit to the function $\xi $ (e.g., Marulli et al. 2012). 

Here we carry out something similar: a weighted fit, 
but as a function of $\Delta z$ and $z\Delta \theta $, instead of $s_\parallel $ and $s_\perp $,
and, in order to solve better the quasi-degeneracy of $\beta $ and $y$, we set a constraint for
$\beta $ to follow $\xi (s,\mu =1)=K(z)R^{-\gamma (z)} f(\mu =1,\beta ,\gamma )$---an amount which is
independent of $y$. We exclude from the fit the pixel (1,1), because it contains higher
nonlinear effects of the correlation and errors from binning. In practice,
instead of taking strictly $\mu =1$ (equivalent to $z\Delta \theta =0$), we
take $z\Delta \theta =0.00125$, i.e., the first column (for the galaxies' samples).
The effect of random velocities or the ``finger of God'' is negligible in the calculation
of this last quantity for our bin sizes of $\Delta (\Delta z)=0.0025$ (for the galaxies' samples): numerical calculations give errors of $\sim 3$\% in $\xi (s,\mu =1)$ with respect
to the inclusion of velocity dispersions in $\Delta z$ with a Gaussian distribution
of $\sigma \sim 0.001$, which is negligible in comparison with other sources of
errors. If we obtained $\beta <0$, we would set $\beta =0$ and recalculate the error bar to get 68\% probability in the range $\beta \ge 0$ within 1$\sigma $.
The error bars of $\gamma $, $y$ and $\beta $ are given by the values which produce a 
$\chi ^2<{\rm Min}[\chi ^2]+3.5$ (the value of 3.5 defines the region with a 68\% confidence level (CL) in a three-free-parameter fit; Avni 1976). Note that the error bars of $\beta $ and $y$ should include both the transmitted errors for
a fixed value of $\gamma $ and the variations of those values due to the variation of $\gamma $ within
the range of its error bar; we will do so throughout this paper.   

\begin{figure}
\vspace{1cm}
{\par\centering \resizebox*{8cm}{8cm}{\includegraphics{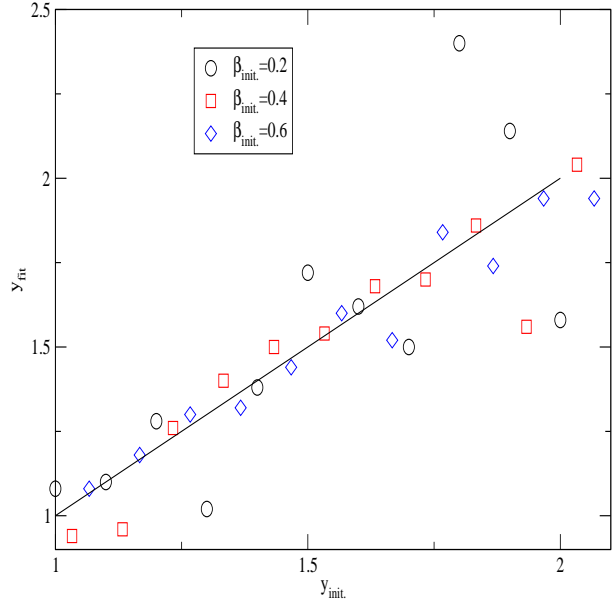}}
\par\centering}
\caption{Monte Carlo simulation with $\gamma _{\rm init.}=1.8$, different values of
$\beta _{\rm init.}$ and noise with relative rms equal to $\frac 
{\Delta \xi }{\xi }(s)=\frac{0.02}{\sqrt{s/0.0025}}
\frac{1+\xi }{\xi }$ with $(\Delta z )_{\rm max}=(z\Delta \theta )_{\rm max}=0.02$.
Solid line stands for $y_{\rm fit}=y_{\rm init.}$.}
\label{Fig:MonteC}
\end{figure}

Tests have been carried out with Monte Carlo simulations and we could see that
we recover approximately the introduced parameters. For instance, in Figure \ref{Fig:MonteC}, we show
the recovered values of $y_{\rm init.}$ versus the introduced values of the fit $y_{\rm fit}$ 
in a Monte Carlo simulation. The conditions are $\gamma _{\rm init.}=1.8$, different values of
$\beta _{\rm init.}$, and noise with relative rms equal to $\frac 
{\Delta \xi }{\xi }(s)=\frac{0.02}{\sqrt{s/0.0025}}
\frac{1+\xi }{\xi }$ with $(\Delta z )_{\rm max}=(z\Delta \theta )_{\rm max}=0.02$ over an area 
subdivided into 8x8 bins of size 0.0025. 
As can be observed, we recover roughly the value of $y$ plus some noise.

Many authors have already realized on the advantages of using the anisotropic correlation function to measure 
the ratio of observed angular size to radial/redshift size and
applied it as a possible way to constrain cosmological models.
It was analyzed in Ballinger et al. (1996), Marulli et al. (2012), Phillips (1994) for quasar pairs, Ryden (1995) for voids, Kim \& Croft (2007) for clusters, and Nusser (2005) for radio emission in the epoch of reionization, among others. It was applied to real data of several surveys with galaxies (Blake et al. 2011), luminous red galaxies (Ross et al. 2007; Okumura et al. 2008; Chuang \& Wang 2012; Reid et al. 2012), Lyman-break galaxies (Da \^Angela et al. 2005b), QSOs (Hoyle et al. 2002; Outram et al. 2004; Da \^Angela et al. 2005a; Ivashchenko et al. 2010), galaxy pairs (Marinoni \& Buzzi 2010), cross-correlation of QSOs and luminous red galaxies (Mountrichas et al. 2009a), cross-correlation of galaxies and cluster of galaxies (Mountrichas et al. 2009b), and voids (Sutter et al. 2012), among others.
Most of these authors analyzed the correlation on small scales ($\lesssim 50$ $h^{1}$Mpc) and very few of them (Okumura et al. 2008; Chuang \& Wang 2012) use larger scales. 


\section{Calculation of the anisotropic two-point correlation function}
\label{.2pcf_calc}

Let us suppose we have a survey completely covering some area with some homogeneous criterion. The method to calculate the anisotropic two-point correlation function
is as follows.
We generate another mock survey in the same area with a random (Poisson) distribution of galaxies and with the same distribution of redshifts as the data.  
Then we use a generalization of the Landy \& Szalay (1993) algorithm to get 
the anisotropic two-point correlation function,

\begin{equation}
\xi (\Delta z, z\Delta \theta ;z)=1+\frac{{\rm DD}(\Delta z, z\Delta \theta ;z)}{f_{\rm RR} {\rm RR}(\Delta z, z\Delta \theta ;z)}
\end{equation}\[
-\frac{f_{\rm DR} {\rm DR}(\Delta z, z\Delta \theta ;z)}{f_{\rm RR} {\rm RR}(\Delta z, z\Delta \theta ;z)}-\frac{f_{\rm RD} {\rm RD}(\Delta z, z\Delta \theta ;z)}{f_{\rm RR} {\rm RR}(\Delta z, z\Delta \theta ;z)}
,\]
where DD, RR, DR and RD are the number of data--data, random--random, data--random and random--data pairs, and
$f_{XY}$ denotes their normalization to data--data, i.e., $f_{XY}XY=\frac{N_{D,1}(N_D-1)}{N_{X,1}(N_Y-1)}XY$; $N_D$ and $N_R$ are the total number of sources in the data and in the random sample;  
$N_{D,1}$ and $N_{R,1}$ are the total number of sources in the data and in the random sample which contribute as first object in the pair for a given $z$ (within some interval of redshifts). Notice that $N_{D,1}<N_D$ and
$N_{R,1}<N_R$ because we remove as first object of the pair the sources which are within an angular distance 
from a border of the survey lower than $\theta _{\rm max}(z)$.
Consequently, DR is slightly different from RD: 
the parent galaxies (the first character $D$ or $R$) are limited to 
a zone away from the borders of the selected area, whereas the second character includes all the galaxies
of the catalog within a given distance from the parent source.
In our case, we take a number of random points equal to the number of sources in the data. However,
the result of the algorithm is independent of that number, provided that it is at least as large as the number
of data points (Sylos Labini et al. 2009) and it is high enough to avoid the production of further noise due to sampling errors. The relative error of $(1+\xi )$ is equal to the inverse of the root square of the number of pairs (Betancort-Rijo 1991; Ross et al. 2007) but, since we manage a huge number of pairs both in the real data and in the random sample, this error is small in comparison to other errors. This number must not be very high in order to have a reasonable computation time to calculate the number of pairs. Other estimators of the two-point correlation function give very similar results even at large scales (e.g., Landy \& Szalay 1993; Porciani \& Norberg 2006; Ross et al. 2007; Sylos Labini et al. 2009).

The error of $\xi (\Delta z, z\Delta \theta ;z)$ is evaluated through a field-to-field method (Ross et al. 2007, Section 2.4):
\begin{equation}
\label{errxi}
\sigma _{\xi }=\sqrt{\frac{1}{N_f-1}\sum _{i=1}^{N_f}\frac{{\rm DR}_i}
{\rm DR}[\xi _i-\xi ]^2}
,\end{equation}
where $N_f$ is the total number of subsamples (fields), and the index $i$ indicates the statistics within the
subsample $i$. In our case, we take by default $N_f=10$, dividing the covered region of the sky into 10 subregions. 

Although Landy \& Szalay (1993) claim that the error given by their estimator should be nearly Poissonian, the Poissonian error significantly underestimates the total error at large distances (Ross et al. 2007,  Section 2.4). The jackknife or bootstrap methods provide errors of the same order (Ross et al. 2007, Section 2.4); however, if there were important large-scale variations of the correlation, the jackknife or bootstrap algorithms would underestimate the total rms measured in the field-to-field algorithm 
(L\'opez-Corredoira et al. 2010).
For point distributions, there is a component of the total error that is caused by the finiteness of the number of points (Betancort-Rijo 1991) and is closely related to that given by the resampling techniques such as jackknife or bootstrap. The field-to-field fluctuations also include intrinsic fluctuations in the large-scale structure of galaxies. This technique of field-to-field is equivalent to carrying out several independent realizations, one for each of the subsamples in which is divided the whole, equivalent to analyzing the rms in multiple mock catalogs generated randomly, with the advantage that here we can be pretty sure that the generated subsamples obey perfectly the statistical distribution of the real sky since they are part of the whole real sky, whereas mock catalogs may depend on how the algorithm generates the samples.

One may wonder whether the correlations between the different subsamples can change the rms of
Equation (\ref{errxi}) with respect to the real one, i.e., whether the terms of the covariance matrix are significant. The answer is that this effect is negligible provided that $\frac{4}{\pi ^3}\Delta \theta \ll 1$
(see Appendix \ref{.2subsamples_tpcf}),
which is true in our case.

Therefore, the field-to-field algorithm seems an approximate method to take all the errors into account
in a simple and direct way.

\subsection{Application to samples with mean density depending on redshift}
\label{.vardens}

Using all available galaxies in a catalog would be better in order to gain higher signal-to-noise in the statistics, provided that the criterion of selection of sources does not vary with the position. In principle, this is possible for cosmological purposes, since cosmological parameters are 
totally independent of the evolution of galaxies or their clustering. The function $y(z)$ will only depend on the cosmological model. 
Also, $y(z)$ does not depend on the kind of sources examined or the wavelength; we can even mix different types of sources for different $z$ without $k$-corrections, so the selection effects (Malmquist bias and others) will not have any influence on the result. In a limiting-magnitude survey covering some area, we will have more luminous sources at higher redshifts, different observed ``at-rest'' wavelengths, etc., but it does not matter in the measurement of the geometry of the universe. Nonetheless, one must take into account the effect of a variable $\langle \rho \rangle (z)$ in the calculation of the two-point correlation function.

If the maximum comoving radius, $s$, were very small, and the galaxies within a sphere of radius $s_{\rm max}$ were similar in their properties, $\xi $ might be calculated as usual in a homogeneous space. But in our case, we must consider some effect of variation for larger $s$ due to the variability of $\langle \rho \rangle (z)$ within the considered effects, mainly because of the Malmquist bias. This variability of the mean density produces a
contribution to the two-point correlation function, which must be subtracted in order to see only the contribution from the clustering of galaxies. This
correction is

\begin{equation}
\xi ^{\rm corrected}(\Delta z, z\Delta \theta ;z)=-1+\frac{1+\xi (\Delta z, z\Delta \theta ;z)}{1+\xi ^{0}(\Delta z, z\Delta \theta ;z)}
,\end{equation}
where $\xi $ is the measured correlation, and $\xi ^0$ is the correlation given by a random (i.e., uncorrelated, locally Poisson) distribution of sources within a field $\langle \rho \rangle (z)$. In Appendix \ref{.varrho}, we demonstrate that, neglecting terms of $(\Delta z)^n$ with $n\ge 4$,
\begin{equation}
\xi ^{0}(\Delta z, z\Delta \theta ;z)\approx \frac{1}{2}\frac{\rho ''(z)}{\rho (z)}(\Delta z)^2
.\end{equation}

This correction will be applied to our calculations.

\section{Application of the method to the data}
\label{.application}

\subsection{Spectroscopic survey of galaxies from SDSS-III/BOSS}
\label{.boss}

The Sloan Digital Sky Survey III (SDSS-III)/Baryon Oscillation Spectroscopic Survey (BOSS; Eisenstein et al. 2011; Ahn et al. 2012; Bolton et al. 2012; Smee et al. 2013; Dawson et al. 2013) will target, when finished, 1.5 million galaxies in ten thousand square degrees, $i < 19.9$, selected in color--magnitude space to be high-luminosity systems at large distances. The selection criteria are the union of two cuts designed to select targets in two different redshift intervals. Cut I, aimed at the interval $0.2 < z < 0.4$, is defined by $r < 13.6 + c_{||}/0.3$, $|c_{\perp }|<0.2$, and 
$16 < r < 19.5$. Cut II, aimed at redshift $z > 0.4$, is defined by $d_{\perp } > 0.55$, $i < 19.86+1.6 \times (d_{\perp } - 0.8)$, and $17.5 < i < 19.9$, where the colors $c_{||}$, $c_\perp$, and $d_\perp $ are defined to track a stellar population passively evolving with redshift: $c_{||} = 0.7 \times (g-r) + 1.2 \times (r-i-0.18)$, $c_{\perp } = (r-i) - (g-r)/4 - 0.18 $, and $d_{\perp } = (r-i) - (g-r)/8$.
These selection criteria produce a roughly constant comoving space density $3\times 10{^{-4}}$ $h^3$Mpc$^{-3}$ to $z = 0.6$ with a slight peak at $z\approx 0.55$ followed by a declining space density to $z\approx 0.8$.
The average accuracy in the redshift determination of these sources is $\Delta z\approx 0.00013$ (Dawson et al. 2013). The survey has started in the fall of 2009 and is planned to be finished in approximately 2014 July. At present, we use the BOSS Data Release 10 version 5 to do statistical analyses of large-scale structure (Anderson et al. 2012) with uniform coverage. Some masks were applied to avoid regions with some saturated sources or low completeness (Anderson et al. 2012). This subsample contains 830,089 galaxies.

In this sample, we use the galaxies within $0.13\le z\le 0.77$ (793,573 galaxies).
Given the considerations in Section \ref{.vardens}, we do not need a homogeneous sample with respect to the
variation in $z$. Some Malmquist bias is expected to be introduced, even if the sources were selected to be kept
almost constant for some range of redshifts, but our procedure can use samples with variable $\langle \rho \rangle (z)$. Nonetheless, in order to assure homogeneity in the angular scale and avoid border effects, we select as first object in pairs which contribute to the two-point correlation function only those of them whose circles of angular radii $\theta _{\rm max}=0.02/z$ radians are totally covered; indeed, for more security, we require them to be totally covered within a distance of 
$\theta _{\rm max}+0^\circ .3$ in the SDSS sample.
Thus, the number of parent galaxies (first one of the galaxies in a pair for the correlation calculation) is 456,451. These are divided into three redshift
bins, as is described in Table \ref{Tab:ybg}.

We also take into account that the minimum separation between fibers is sep$=62''=3.0\times 10^{-4}$ radians, so we must add $\frac{1}{2}{\rm sep}\ z$ to the value of the mean $z\Delta \theta $ for the lowest $\theta $ bin, which is negligible in our scales.

Previous analyses of the anisotropic correlation function with previous releases of BOSS data were also carried out by White et al. (2011) and Reid et al. (2012), although with different scales, methods and purposes from the present paper.

\begin{landscape}
\begin{table*}
\caption{Values of $y$, $\beta $, and $\gamma $ Obtained from Our Data. First three rows correspond to galaxies from BOSS, the following two rows correspond to QSOs from SDSS-II, and the last
two rows correspond to QSOs from BOSS. $\chi _{\rm red.}^2$ corresponds to the reduced chi-square of
the fit in the 8x8 pixels excluding the pixel (1,1).}
\begin{center}
\begin{tabular}{cccccccc}
\label{Tab:ybg}
Redshift range & $(\Delta z)_{\rm max}=(z\Delta \theta )_{\rm max}$ & \# parent sources & $\langle z\rangle $ & $\chi _{\rm red.}^2$ & $\gamma $ & $\beta $ & $y$ \\ \hline
$[0.15-0.35)$ & 0.02 & 50725 & 0.289 & 0.44 & $1.50\pm 0.07$ & $0.25\pm 0.21$ & $1.12\pm 0.27$ \\ 
$[0.35-0.55)$ & 0.02 & 240745 & 0.476 & 10.1 & $1.98\pm 0.07$ & $0.17\pm 0.10$ & $1.26\pm 0.23$ \\
$[0.55-0.75)$ & 0.02 & 177150 & 0.615 & 2.93 & $2.06\pm 0.06$ & $0.13\pm 0.10$ & $1.14\pm 0.20$ \\ \hline
$[0.75-1.75)$ & 0.08 & 23972 & 1.298 & 0.72 & $0.54\pm 0.12$ & $0^{+0.46}_{-0}$ & $1.40\pm 1.03$ \\
$[1.75-2.75)$ & 0.08 & 12169 & 2.056 & 0.75 & $0.60\pm 0.14$ & $0^{+0.34}_{-0}$ & $5.80\pm 3.72$ \\ \hline
$[2.00-2.75)$ & 0.08 & 20936 & 2.436 & 0.91 & $0.62\pm 0.12$ & $0^{+0.60}_{-0}$ & $1.88\pm 1.13$ \\
$[2.75-3.50)$ & 0.08 & 10305 & 3.056 & 0.42 & $0.50\pm 0.12$ & $0^{+0.49}_{-0}$ & $7.06\pm 9.80$ \\ \hline
\end{tabular}
\end{center}
\end{table*}
\end{landscape}

\subsection{Spectroscopic surveys of QSOs from SDSS-II and SDSS-III/BOSS}
\label{.qsoSDSS}

We also use two catalogs of QSOs, both from the SDSS-III/BOSS
and SDSS-II, respectively. The first one is deeper but more constrained
in area and redshift range, so the application to SDSS-II QSOs is interesting
too. 

The SDSS-II quasar catalog (DR7 quasar catalog; Schneider et al. 2010) contains 105,783 spectroscopically confirmed quasars. The catalog consists of the 
SDSS\footnote{Details on the general characteristics of the SDSS survey, its telescope, camera, and others, can be found in Gunn et al. (1998, 2006) and York et al. (2000).} objects that have luminosities larger than $M_i=-22.0$ (in a cosmology with $H_0=70$ km s$^{-1}$Mpc$^{-1}$, $\Omega_m= 0.3$, and $\Omega_\Lambda=0.7$),
and have at least one emission line with FWHM larger than 1000 km s$^{-1}$ or have interesting/complex absorption features; they are fainter than $i>15.0$. The catalog covers an area of 9380 deg$^2$. The quasar redshifts range from 0.065 to 5.46, with a median value of 1.49, and typical quoted redshift errors of $\approx 0.004$. The catalog was created by inspecting all spectra that were either targeted as quasar candidates or classified as a quasar by the spectroscopic pipelines, so it provides the most reliable classifications and redshifts of SDSS quasars.
We select the sources with the flag ``UNIFORM\_TARGET=1'', which constitutes a statistical sample appropriate for clustering studies: 59,514 sources. We only use the galaxies within $0.67\le z\le 2.83$. From these, we select the parent sources whose surrounding area is totally covered within a radius of 
$\theta _{\rm max}=5.2\times 10^{-4}+0.08/z$ radians;
thus the number of parent QSOs is 36,141. These are divided into two redshift bins, as is described in Table \ref{Tab:ybg}.
We also take into account that the minimum separation between fibers is sep$=55''=2.7\times 10^{-4}$ radians, so we must add $\frac{1}{2}{\rm sep}\ z$ to the value of the mean $z\Delta \theta $ for the lowest $\theta $ bin, which is negligible in our scales.
Previous analyses of the anisotropic correlation function with previous releases of SDSS-II QSO data were also carried out by Ivashchenko et al. (2010), although with different scales, methods, and purposes of the present paper.

The SDSS-III/BOSS quasar catalog (P\^aris et al. 2012) includes all BOSS objects that were targeted as quasar candidates during the survey, are spectrocospically confirmed as quasars via visual inspection, have luminosities $M_i[z=2]<-20.5$ (in a $\Lambda$CDM cosmology with $H_0=70$ km s$^{-1}$Mpc$^{-1}$, $\Omega_{\rm m}$=0.3, and $\Omega_{\Lambda}$= 0.7), and either display at least one emission line with a FWHM larger than 500 km s$^{-1}$ or, if not, have interesting/complex absorption features. Their redshifts were checked by a visual inspection, which provides a reliable and secure redshift estimate for each quasar, with a minimum error of $\approx 0.0017$.
The average error in the redshift determination of the sources is $\approx 0.003(1+z)$ (White et al. 2012).
Around 10\% of QSOs were included in the spectroscopic catalog of SDSS-II and 90\% are new, so we can consider its results to be almost independent from SDSS-II results. The number of quasars with $z\ge 2.15$ is much higher than in SDSS-II, so this survey has the advantage of allowing a better exploration of high-redshift large-scale structure.
At present, we use the version DR10Q\_alpha\_3 (P\^aris et al. 2013). We select the point-like objects with SPECPRIMARY=1 that belong to the uniform QSO\_CORE\_MAIN sample, thus providing a statistical sample appropriate for clustering studies (White et al. 2012). The total number of QSOs is 80,380. We use only the sources within $1.92\le z\le 3.58$. From these, we use as parent sources those QSOs that are totally covered within a radius $\theta _{\rm max}=5.2\times 10^{-4}+0.08/z$ radians and have a pipeline redshift equal to the visual inspection redshift within 3-$\sigma $; thus the number of parent QSOs is 31,241. These are divided into two redshift bins, as described in Table \ref{Tab:ybg}.
Again, we take into account that the minimum separation between fibers is sep=62'', as explained previously.

\subsection{Two-point correlation functions and parameters obtained from them}

\begin{figure*}[htb]
\vspace{1.2cm}
{\par\centering \resizebox*{6cm}{16.1cm}{\includegraphics{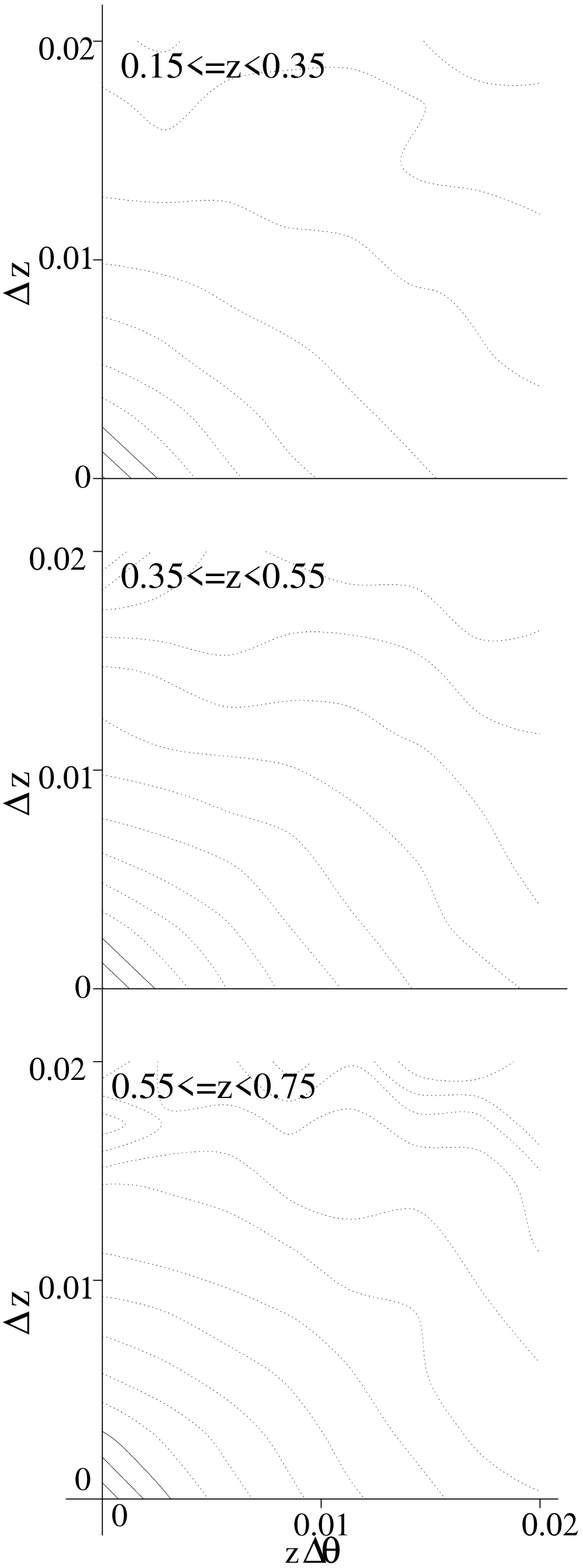}}
\hspace{1cm}
\resizebox*{6cm}{16.1cm}{\includegraphics{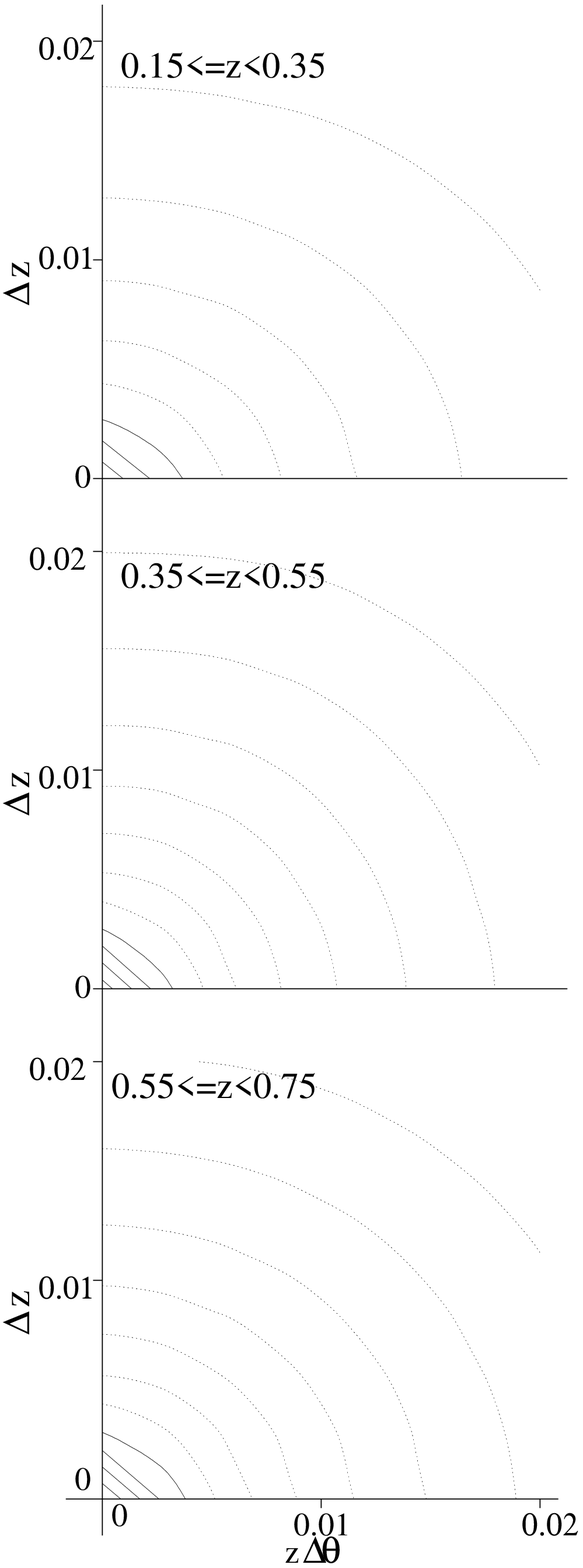}}
\par\centering}
\caption{Left: $\log_{10}\xi (\Delta z, z\Delta \theta )$ for the galaxies of SDSS-III/BOSS at different redshift ranges. Right: best fit of $\log_{10}\xi (\Delta z, z\Delta \theta )$
with the parameters given in Table \protect{\ref{Tab:ybg}}.
Solid lines stand for null or positive values, whereas dotted lines stand for negative values; 
the step of each contour is 0.2 dex. The pixel binning is $0.0025\times 0.0025$.}
\label{Fig:elipses3}
\end{figure*}

\begin{figure*}[htb]
\vspace{1.2cm}
{\par\centering \resizebox*{6cm}{10.7cm}{\includegraphics{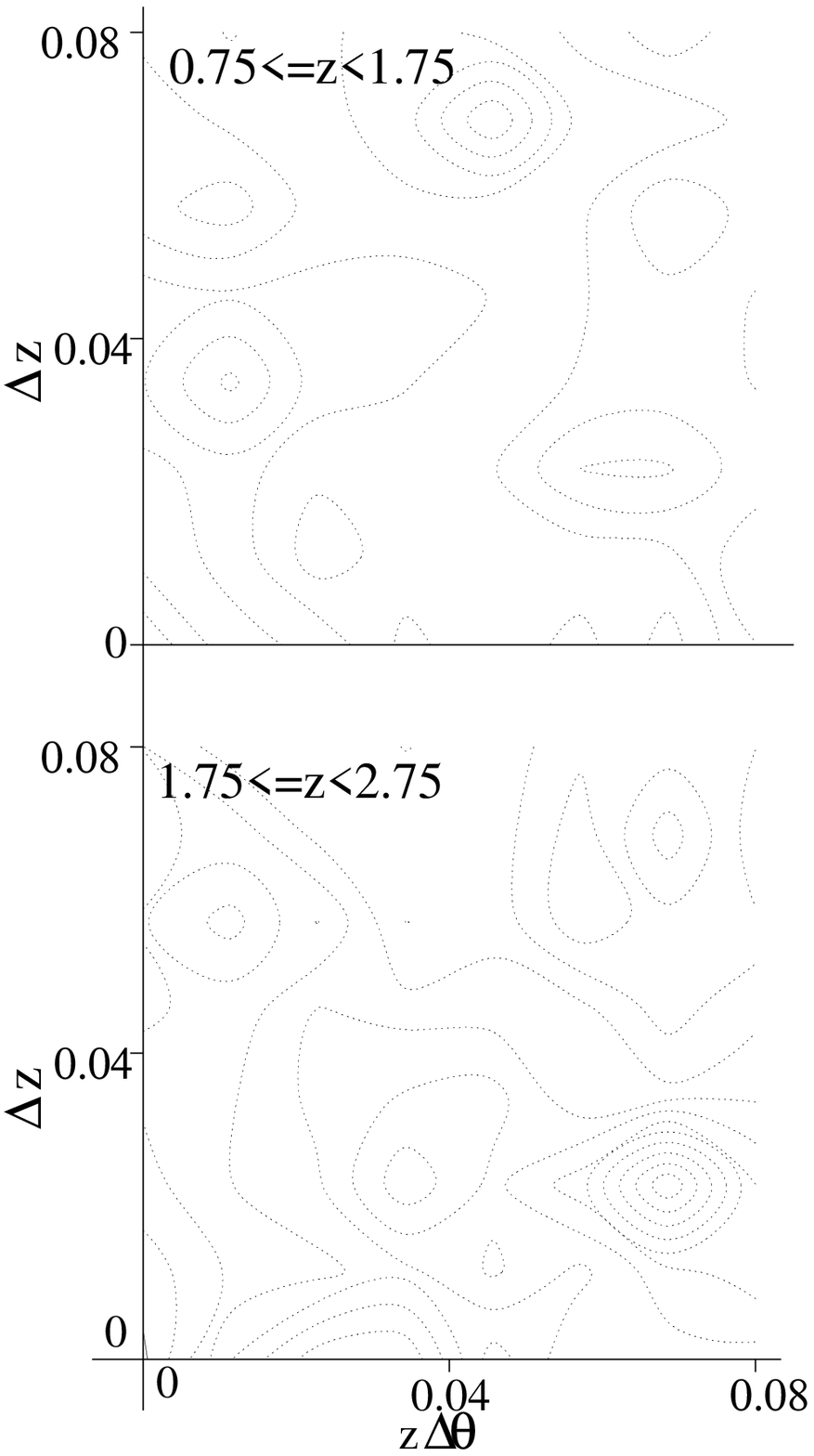}}
\hspace{1.cm}
\resizebox*{6cm}{10.7cm}{\includegraphics{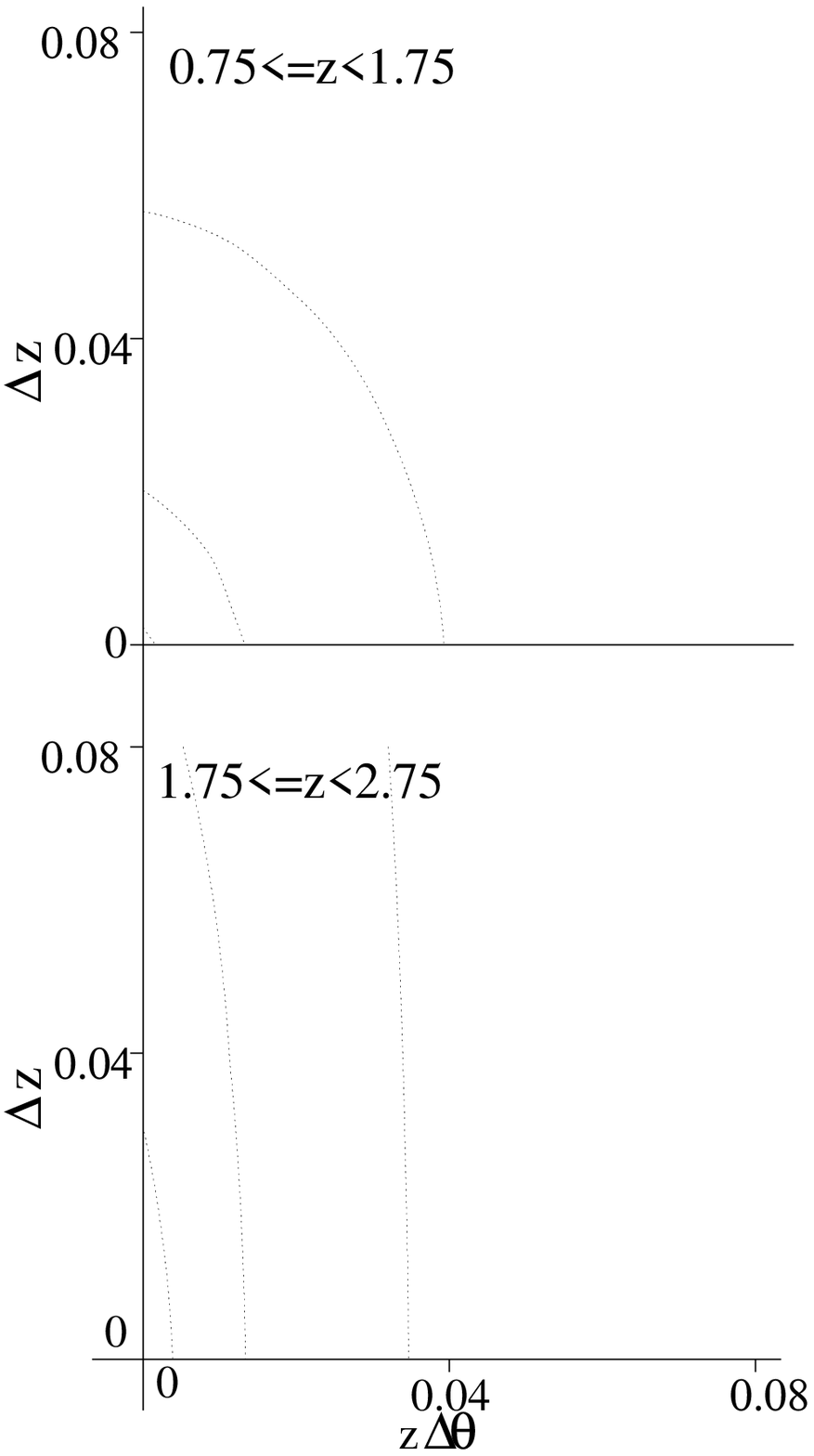}}
\par\centering}
\caption{Left: $\log_{10}\xi (\Delta z, z\Delta \theta )$ for the QSOs of SDSS-II at different redshift ranges. Right: best fit of $\log_{10}\xi (\Delta z, z\Delta \theta )$
with the parameters given in Table \protect{\ref{Tab:ybg}}.
All lines (dotted) represent negative values; 
the step of each contour is 0.2 dex. The pixel binning is $0.010\times 0.010$.}
\label{Fig:elipses2Q}
\end{figure*}

\begin{figure*}[htb]
\vspace{1.2cm}
{\par\centering \resizebox*{6cm}{10.7cm}{\includegraphics{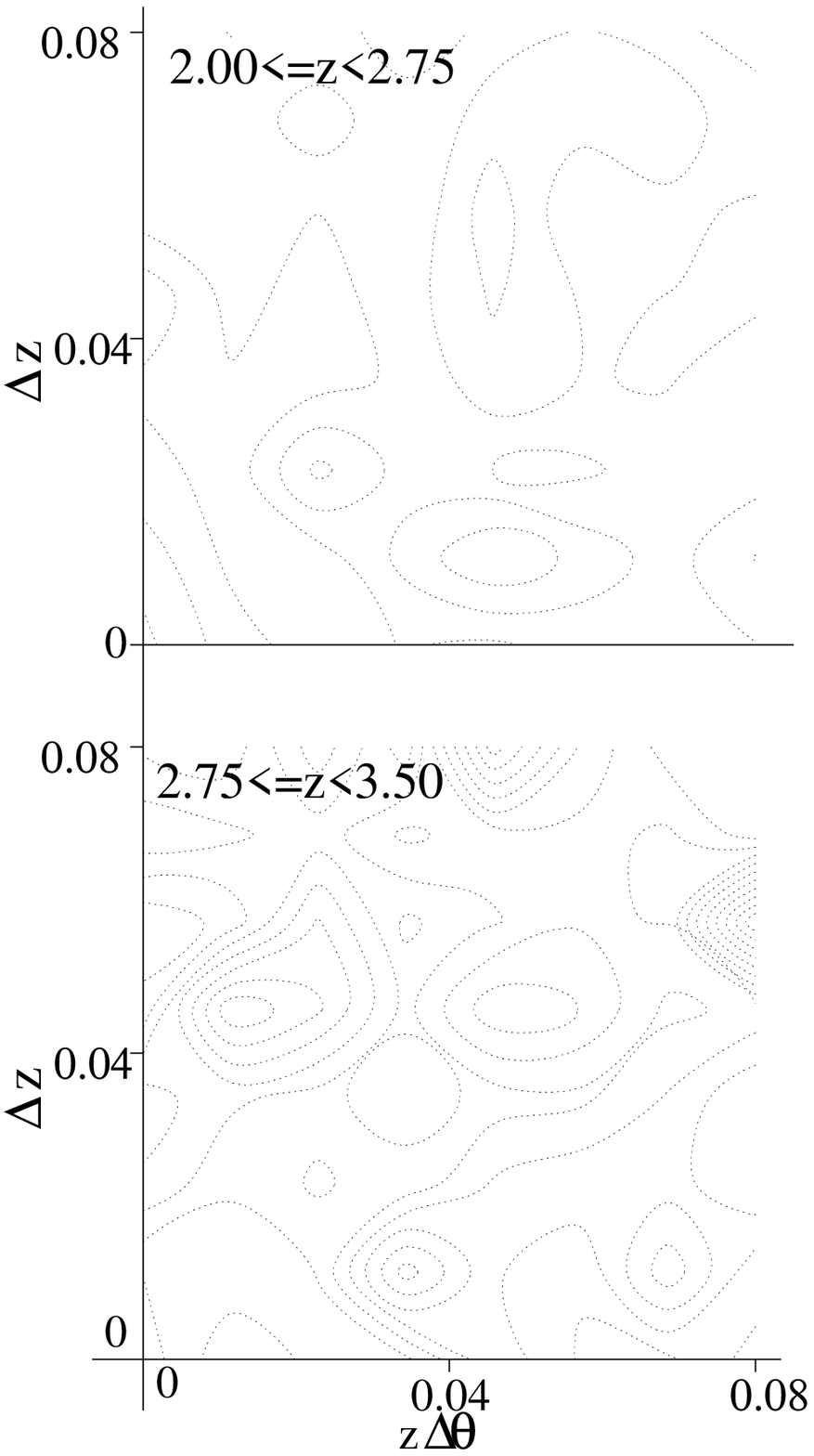}}
\hspace{1cm}
\resizebox*{6cm}{10.7cm}{\includegraphics{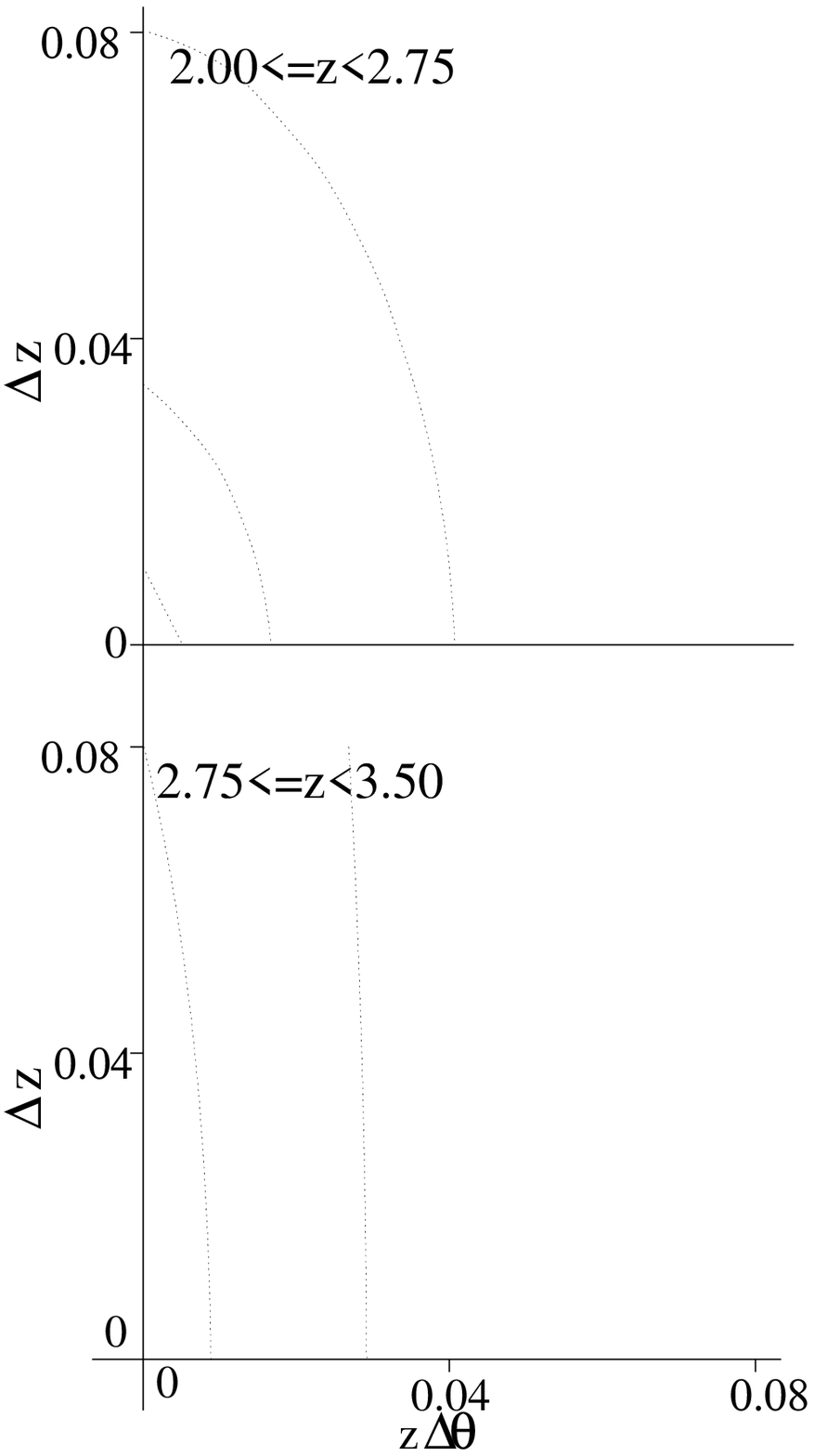}}
\par\centering}
\caption{Left: $\log_{10}\xi (\Delta z, z\Delta \theta )$ for the QSOs of SDSS-III/BOSS at different redshift ranges. Right: best fit of $\log_{10}\xi (\Delta z, z\Delta \theta )$
with the parameters given in Table \protect{\ref{Tab:ybg}}.
All lines (dotted) represent negative values; 
the step of each contour is 0.2 dex. The pixel binning is $0.010\times 0.010$.}
\label{Fig:elipses2Q2}
\end{figure*}

The application of the Landy \& Szalay algorithm to our SDSS-III/BOSS galaxies gives the anisotropic two-point correlation function that is shown in Figure \ref{Fig:elipses3} for different redshift ranges of width 0.2 for $(\Delta z)_{\rm max}=(z\Delta \theta )_{\rm max}=0.02$; this is equivalent to
45.8 and 52.9 $h^{-1}$Mpc along the line-of-sight and perpendicular directions, respectively, at $z=0.5$ (assuming a standard concordance model with $\Omega _m=0.3$). For the QSOs of SDSS-II or BOSS, whose two-point correlation functions are plotted in Figures \ref{Fig:elipses2Q} and \ref{Fig:elipses2Q2}, we take a larger scale because there are very few QSOs with small distances and also the redshift errors are larger. Here we use bins in redshift of width 0.75 and $(\Delta z)_{\rm max}=(z\Delta \theta )_{\rm max}=0.08$, which is equivalent to 80.8 and 145.0 $h^{-1}$Mpc along the line-of-sight and perpendicular directions respectively at $z=2.0$ (assuming a standard concordance model with $\Omega _m=0.3$). The values of the reduced chi-Square of the fit,
$\chi _{\rm red.}^2$, given in Table \ref{Tab:ybg}, correspond to the 8x8 pixel analysis excluding the pixel (1,1); i.e., in 63 pixels and with three degrees of freedom.

\begin{figure}[htb]
\vspace{1.2cm}
{\par\centering \resizebox*{8cm}{8cm}{\includegraphics{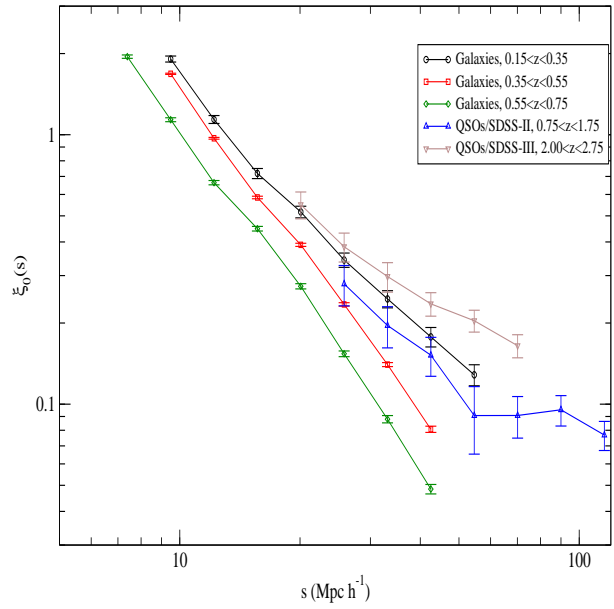}}
\par\centering}
\caption{Average isotropic two-point correlation functions, $\xi _0(s)$, 
assuming the standard concordance model (model 1 of Appendix 
\ref{.cosmomodels}) and the values of $y$ given in Table \protect{\ref{Tab:ybg}}.}
\label{Fig:xi0}
\end{figure}

In Figure \ref{Fig:xi0}, we plot the average isotropic two-point correlation functions, $\xi _0(s)\equiv \int _0^1d\mu \,\xi(s,\mu )$, assuming the standard concordance model (model 1 of Appendix 
\ref{.cosmomodels}) and the values of $y$ given by Table \ref{Tab:ybg}. As can be observed
in this log--log plot, the functions are well represented by straight lines within the error bars, which means that the assumption of a power law for the two-point correlation function is approximately correct. 
The minimum scale used for the fits is the pixel size, which is in all cases $\frac{1}{8}(\Delta z)_{\rm max}$. We exclude the data in the fifth and the seventh rows of Table \ref{Tab:ybg} because their 
$y$ values are very large; consequently there is a very small range of distances between the minimum $s$ that satisfies this condition
and the maximum $s$. 
The average deviation from a power-law fit is $\approx 5$\% for the three plots of galaxies, and $\approx 8$\% for the QSOs. However, the error bars are very small for the galaxy samples (smaller than 0.01 dex in some bins), so we can detect in them a significant departure from a power law. The respective $\chi _{\rm red}^2$ for the five weighted linear fits in the log--log of $\xi _0(s)$ given in Figure \ref{Fig:xi0} are 3.6, 14.5, and 7.4 for the three sets with galaxies and 0.84 and 0.40 for the two sets with QSOs. Certainly, the first three fits for the galaxies are much worse whereas plots of the quasars (the last two) have an acceptable fit with a power-law. Nonetheless, this must not confuse the
reader: while the log--log fits of the galaxies present smaller deviations ($\sim 5$\%), they produce
very high $\chi _{\rm red}^2$ values when combined
with the $\sim 1$\% error bars.
Some slight deviations from a power-law behavior are indeed expected (e.g., Zehavi et al. 2004, although for smaller scales) but their effects are small in comparison with other sources of error in our analysis.

Reid et al. (2012) published the correlation function for BOSS galaxies at $z=0.57$ with similar values of the amplitude to those we obtained in Figure \ref{Fig:elipses3}/down (with average $z=0.615$). Reid et al. (2012) gave an amplitude of $s^2\xi _0(s)\approx 110$ for $s=25\ h^{-1}$ Mpc and $s^2\xi _0(s)\approx 90$ for $s=40\ h^{-1}$ Mpc, whereas we obtain (for the standard cosmology) $s^2\xi _0(s)\approx 100$ for $s=25\ h^{-1}$ Mpc and $s^2\xi _0(s)\approx 88$ for $s=40\ h^{-1}$ Mpc.

\begin{figure}[htb]
\vspace{1.2cm}
{\par\centering \resizebox*{8cm}{8cm}{\includegraphics{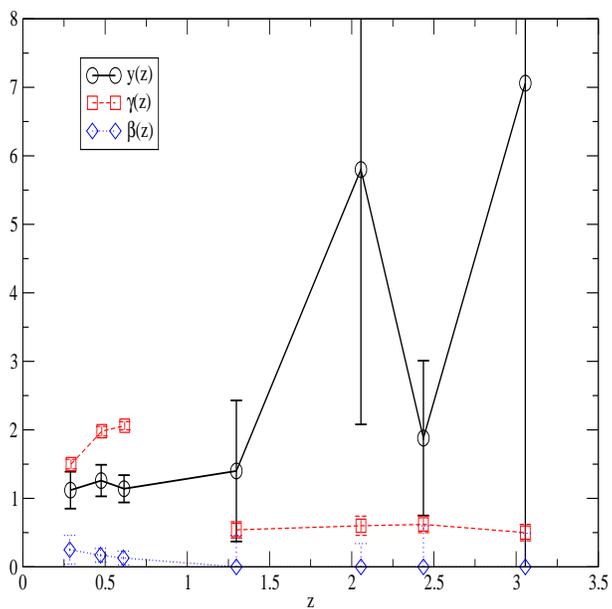}}
\par\centering}
\caption{Values of $y$, $\beta $, and $\gamma $ obtained from our analysis
of BOSS galaxies and SDSS-II QSOs.}
\label{Fig:ybg}
\end{figure}

When we apply the method described in Section \ref{.2pcf} to these
correlation functions, we get the values of $\gamma $, $\beta $, and $y$ that are
listed in Table \ref{Tab:ybg} and plotted in Figure \ref{Fig:ybg}. The values
of $\gamma $ are lower for QSOs because they are observed at larger scales. 
The values of $\beta $ we obtained for the first two bins with BOSS/galaxies are 
roughly compatible with previous values in the literature ($\beta \approx 0.3$), whereas the third bin 
($\beta (z=0.615)=0.13\pm 0.10$) gives a significantly smaller value than, e.g., 
$\beta ({\rm low}\ z)=0.52\pm 0.26$ (Hamilton 1998), $\beta (z=0.55)=0.40\pm 0.05$ (Ross et al. 2007), 
and $\beta (z=0.57)=0.34\pm 0.03$ (Reid et al. 2012). If this were confirmed, it could mean that there is a conspicuous increase of the bias with $z$ for $z>0.5$, while $\Omega _m(z)$ does not change too much (in the interpretation of the standard cosmology). But it is also possible that this result is a $\sim 2\sigma $ statistical negative fluctuation and 
the value of $\beta (z=0.615)\approx 0.3$ is most likely the appropriate one.

\section{Fits of the cosmological models}
\label{.fitcosmo}

\begin{table*}
\caption{Values of $y$ Obtained Directly or Indirectly from $\Omega _m$ (Marked with (*); sSee Explanation in Text) from Different References, Ordered with Increasing Redshifts.}
\begin{center}
\begin{tabular}{cccc}
\label{Tab:yrefs}
Reference & Sources & $\langle z\rangle $ & $y$ \\ \hline
Sutter et al. (2012) & Voids & 0.05 & $0.99\pm 0.05$ \\
Sutter et al. (2012) & Voids & 0.15 & $1.06\pm 0.07$ \\
Blake et al. (2011) & Galaxies & 0.22 & $1.27\pm 0.18$ \\
Sutter et al. (2012) & Voids & 0.28 & $1.04\pm 0.15$ \\ 
Blake et al. (2011) & Galaxies & 0.41 & $1.07\pm 0.17$ \\ 
Ross et al. (2007) & Lum. red gals. & 0.55 & (*) $1.07\pm 0.12$ \\
Blake et al. (2011) & Galaxies & 0.60 & $1.13\pm 0.10$ \\
Blake et al. (2011) & Galaxies & 0.78 & $1.24\pm 0.15$ \\
Marinoni \& Buzzi (2010) & Galaxy pairs & 0.95 & (*) $1.30\pm 0.40$ \\
Outram et al. (2004) & QSOs & 1.40  & (*) $1.50\pm 0.28$ \\ 
Da \^Angela et al. (2005b) & Lyman break gals. & 3.00 & (*) $2.26\pm 0.40$ \\ \hline 
\end{tabular}
\end{center}
\end{table*}

We have obtained $y(z)$ in the previous section (Table \ref{Tab:ybg}). 
In Table \ref{Tab:yrefs}, we add the values of $y(z)$ obtained directly or
indirectly from the previous literature. That is, either the authors of these
references have derived $y(z)$, or they have calculated $\Omega _m(z)$ assuming
the standard model and we calculate its equivalent $y(z)$ through Equation 
(\ref{yz_std}). We have used only the literature in which
we know there is an explicit calculation of $y(z)$ or $\Omega _m(z)$ without making any
a priori assumption on the redshift distortions and in which only the 
Alcock--Paczy\'nski method was used to derive them, without any further information
obtained through other sources of cosmological data. For instance, we have not used
the very accurate measurement of $\Omega _m(z)$ by Okumura et al. (2008), 
because it was derived by marginalizing the dependence with $\sigma _8$ and the
bias $b$. As another example, we have taken the value
of $\Omega _m$ from Ross et al. (2007) described  their Section 4.3, which was derived only 
with the Alcock--Paczy\'nski method, and not
from their Section 4.4, where they introduce further constraints from
assumptions in the clustering evolution on the standard model.
From Da \^Angela et al. (2005b), we take their result without constraints
from linear evolution of density fluctuation.
Note that, in general, our error bars (Table \ref{Tab:ybg}) 
are significantly higher than those from the literature (Table \ref{Tab:yrefs});
the reason is possibly a more conservative account in our 
calculations of the quasi-degeneracy between $\beta $ and $y$.

In total, we have the value of $y(z)$ for 18 different redshifts.
We assume that the data are independent; this is not exactly correct because there may be some correlation among the different catalogs for which the values of $y(z)$ were derived, and there would be a factor in the covariance matrix which
makes the error bars of each pixel not totally independent.
We neglect this factor because the correlation among catalogs is expected to
be small given that each survey is carried out with different selection strategies. Anyway, we do not 
have access to all the catalogs of the different authors in Table \ref{Tab:yrefs} to test it. 
Therefore, our values for the degrees of freedom will be slightly overestimated, but
we accept them as having a correct order of magnitude.

With these data,
we can now calculate the probability of a cosmological model to fit them among
the models given in Appendix \ref{.cosmomodels}. This is carried out by calculating
the reduced $\chi ^2$,
\begin{equation}
\chi _{\rm red}^2=\frac{1}{N-\nu }\sum _{i=1}^N\left(\frac{y(z_i)-y_i^t}{\sigma [y(z_i)]}\right)^2
,\end{equation}
where $\nu $ is the number of free parameters, and $N$ is the total number of data points; hence,
the number of freedom degrees is $N-\nu $.
In our case, $N=18$.  
In models 2, 5, and 6, there are no free parameters. In models 1, 3, and 4, we calculate
the $\chi _{\rm red}^2$ for both the parameters given in Appendix \ref{.cosmomodels}
($\nu =0$) and the following free parameters: 
$\Omega _m$ for models 1 and 3; $\Omega _m$ and $\omega _\Lambda $ for model 1; 
$\Omega _m$ and $\Omega _\Lambda $ (with the constraint
of $\Omega _\Lambda \le 0$ in order to keep a universe with an oscillatory expansion) for model 4. 
Error bars for free parameters are calculated according to the Avni (1976) recipe. 
The results are in Table \ref{Tab:chi2}.
In this table we also include the probability that the model is compatible with the data
by chance, $P=P_{\chi ^2}\times P_{\rm bin}$, where $P_{\chi ^2}$ is the probability given by the $\chi ^2$ test, and $P_{\rm bin}$ is the binomial probability of having the same as or fewer than 
the observed $m$ points with $y\le y_{\rm model}$, i.e.,
\begin{equation}
P_{\rm bin}=1-2^{-N}\left[\sum _{i=m^*+1}^{N-m^*-1}\left (
\begin{array}{c}
N \\
i \\
\end{array}
 \right)\right] 
,\end{equation}\[
m^*\equiv {\rm Minimum}(m,N-m)
.\]
The introduction of the probability $P_{\rm bin}$ is carried out because the
$\chi ^2$ test minimizes the square differences but it does not take into account the sign of these differences, which should be accounted for in a calculation of
the probability. For instance, the model of ``static, linear Hubble law'' with $y_{\rm model}(z)=1\ \forall \ z$ is within the error bars of most of the points, but there is only one point ($m=1$) with $y\le y_{\rm model}$, which is quite unlikely, and this is reflected by the factor $P_{\rm bin}$.

\begin{table*}
\caption{Results of the $\chi ^2$ Test, Best-fit Free parameters (If Any), and
Associated Probability of the models (See text), Using the $N=18$ points of Figure 
\protect{\ref{Fig:models2}}}
\begin{center}
\begin{tabular}{cccccc}
\label{Tab:chi2}
Model & $\nu $ & $\chi _{\rm red,min}^2$ & Free parameters & $m$ & Probability \\ \hline
(1.); $\Omega _m=0.3$, $\omega _\Lambda =-1$ & 0 & 0.27 & --- & 11 & 0.48 \\ 
(1.); $\omega _\Lambda =-1$; $\Omega _m$ free & 1 & 0.26 & $\Omega _m=0.24^{+0.10}_{-0.07}$ & 8 & 0.81 \\ 
(1.); $\Omega _m$, $\omega _\Lambda $ free & 2 & 0.27 & $\Omega _m=0.18^{+0.32}_{-0.16}$, $\omega _\Lambda=-1.2^{+0.7}_{-1.3}$ & 7 & 0.48 \\ 
(2.) & 0 & 1.68 & --- & 15 & $2.7\times 10^{-4}$  \\
(3.); $\Omega _m=0.3$ & 0 & 1.00 & --- & 15 & 0.0034 \\
(3.); $\Omega _m$ free & 1 & 0.73 & $\Omega _m=0^{+0.05}_{-0}$ & 14 & 0.024 \\
(4.); $\Omega _m=1.27$, $\Omega _\Lambda =-0.09$ & 0 & 1.36 & --- & 14 & 0.0044 \\
(4.); $\Omega _m$, $\Omega _\Lambda \le 0$ free & 2 & 1.27 & $\Omega _m=1.22^{+0.12}_{-0.08}$, $\Omega _\Lambda =0^{+0}_{-0.08}$ & 13 & 0.020 \\
(5.) & 0 & 1.46 & --- & 1 & $1.4\times 10^{-5}$ \\
(6.) & 0 & 0.49 & --- & 12 & 0.23 \\ \hline
\end{tabular}
\end{center}
\end{table*}

\begin{table*}
\caption{Results of the $\chi ^2$ Test, Best-fit Free Parameters (If Any), and
Associated Probability of the Models (See Text) with Only the $N=7$ Points
Derived in This Paper}
\begin{center}
\begin{tabular}{cccccc}
\label{Tab:chi2_7}
Model & $\nu $ & $\chi _{\rm red,min}^2$ & Free parameters & $m$ & Probability \\ \hline
(1.); $\Omega _m=0.3$, $\omega _\Lambda =-1$ & 0 & 0.25 & --- & 3 & 0.97 \\ 
(1.); $\omega _\Lambda =-1$; $\Omega _m$ free & 1 & 0.29 & $\Omega _m=0.34^{+0.42}_{-0.24}$ & 3 & 0.94 \\ 
(1.); $\Omega _m$, $\omega _\Lambda $ free & 2 & 0.35 & $\Omega _m=0.31^{+0.69}_{-0.30}$, $\omega _\Lambda=-1.1^{+\infty}_{-\infty}$ & 3 & 0.88 \\ 
(2.) & 0 & 0.53 & --- & 5 & 0.37  \\
(3.); $\Omega _m=0.3$ & 0 & 0.37 & --- & 5 & 0.42 \\
(3.); $\Omega _m$ free & 1 & 0.32 & $\Omega _m=0^{+0.63}_{-0}$ & 4 & 0.91 \\
(4.); $\Omega _m=1.27$, $\Omega _\Lambda =-0.09$ & 0 & 0.46 & --- & 5 & 0.39 \\
(4.); $\Omega _m$, $\Omega _\Lambda \le 0$ free & 2 & 0.56 & $\Omega _m=1.26^{+1.22}_{-0.48}$, $\Omega _\Lambda =0^{+0}_{-0.88}$ & 4 & 0.73 \\
(5.) & 0 & 0.68 & --- & 0 & 0.011 \\
(6.) & 0 & 0.28 & --- & 3 & 0.96 \\ \hline
\end{tabular}
\end{center}
\end{table*}

\begin{table*}
\caption{Results of the $\chi ^2$ Test, Best-fit Free Parameter (If Any), and
Associated Probability of the Models (See Text) with Only the $N=11$ Points
Obtained from the Literature.}
\begin{center}
\begin{tabular}{cccccc}
\label{Tab:chi2_11}
Model & $\nu $ & $\chi _{\rm red,min}^2$ & Free parameters & $m$ & Probability \\ \hline
(1.); $\Omega _m=0.3$, $\omega _\Lambda =-1$ & 0 & 0.27 & --- & 8 & 0.22 \\ 
(1.); $\omega _\Lambda =-1$; $\Omega _m$ free & 1 & 0.29 & $\Omega _m=0.23^{+0.10}_{-0.08}$ & 5 & 0.990 \\ 
(1.); $\Omega _m$, $\omega _\Lambda $ free & 2 & 0.26 & $\Omega _m=0.16^{+0.35}_{-0.14}$, $\omega _\Lambda=-1.22^{+0.74}_{-1.48}$ & 5 & 0.985 \\ 
(2.) & 0 & 2.41 & --- & 10 & $6.3\times 10^{-5}$  \\
(3.); $\Omega _m=0.3$ & 0 & 1.41 & --- & 10 & 0.0019 \\
(3.); $\Omega _m$ free & 1 & 0.93 & $\Omega _m=0^{+0.06}_{-0}$ & 10 & 0.0059 \\
(4.); $\Omega _m=1.27$, $\Omega _\Lambda =-0.09$ & 0 & 2.06 & --- & 9 & 0.0013 \\
(4.); $\Omega _m$, $\Omega _\Lambda \le 0$ free & 2 & 1.63 & $\Omega _m=1.20^{+0.15}_{-0.05}$, $\Omega _\Lambda =0^{+0}_{-0.11}$ & 9 & 0.0026 \\
(5.) & 0 & 1.96 & --- & 1 & $3.3\times 10^{-4}$ \\
(6.) & 0 & 0.63 & --- & 9 & 0.052 \\ \hline
\end{tabular}
\end{center}
\end{table*}

\begin{figure}[htb]
\vspace{1cm}
{\par\centering \resizebox*{8cm}{8cm}{\includegraphics{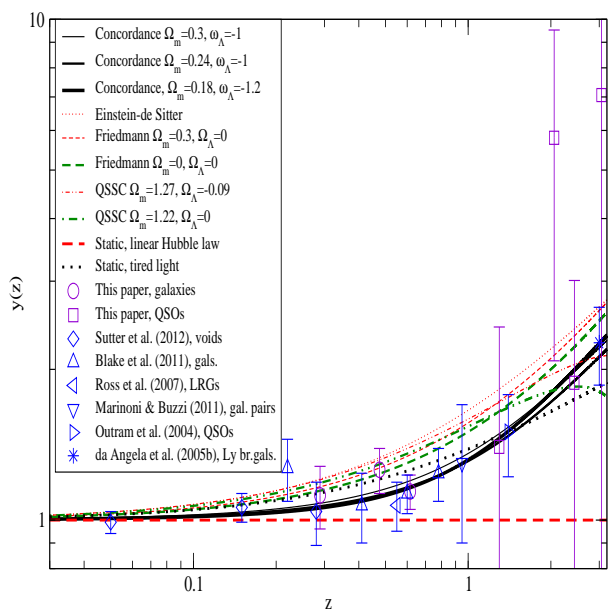}}
\par\centering}
\caption{Log--log plot of the values of $y$ obtained from our analyses
of BOSS galaxies and QSOs and SDSS-II QSOs, plus other values from the literature and the predictions of different models. Green lines indicate that the model is excluded at a $>95$\% CL; red lines indicate that the model is excluded at a $>99$\% CL.}
\label{Fig:models2}
\end{figure}

We observe in Table \ref{Tab:chi2} that some $\chi _{\rm red}^2$ are much lower than unity for $N=18$, more than would be expected statistically. This indicates two things: (1) the previously mentioned fact that the 18 points are not totally independent, and (2) the well-known fact that the scientific community tends to publish results with a dispersion much lower than expected statistically from the error bars, which means that either the error bars were overestimated, or there is a bias in the publication of results towards the preferred value (Croft \& Dailey 2011).
We also find some probabilities which are lower than 0.05
(excluded at a 95\% CL) or even lower than 0.01 (excluded at a 99\% CL).
In Figure \ref{Fig:models2}, we plot all the values of $y(z)$, including our values and those ones from the literature, as well as the different cosmological models.
In Tables \ref{Tab:chi2_7} and \ref{Tab:chi2_11}, we also show the same analysis
with the subsamples of data from our $N=7$ points (Table \ref{Tab:ybg}) and separately the analysis derived from the $N=11$ points from the literature
(Table \ref{Tab:yrefs}). We see that the probabilities are very low for some cases
in the subsamples of 11 points or the whole sample of 18 points; the subsample of 7 points
is not conclusive but it shows the same trends in the probabilities, which
indicates that the exclusion of some cosmological models is robust.

\section{Discussion and conclusions}
\label{.conclusions}

The application of the Alcock--Paczy\'nski test using the anisotropic two-point correlation function was shown here to be a useful tool for discriminating among different cosmological models without any a priori assumptions. We have seen that the
major caveat is the disentanglement with the redshift distortions produced by the peculiar velocities of gravitational
infall, but it is still possible to separate both effects. The error bars are large for the determination of the ratio
of observed angular size to radial/redshift size, $y(z)$; however, with large collections of data from many surveys, it is possible to get some constraints in cosmological models.

We have applied our method to three surveys within SDSS: galaxies from SDSS-III/BOSS, and QSOs from both SDSS-II and SDSS-III/BOSS. 
We have not used the survey of galaxies from SDSS-II because most of their galaxies are at $z\lesssim 0.3$, and the condition of completeness within a radius $\theta _{\rm max}=0.02/z$ would reduce enormously the sample of parent galaxies, producing a result much poorer than with BOSS galaxies. The alternative of exploring smaller distances in the correlation would alleviate this constraint of completeness areas; however, it would introduce the problem of the ``fingers of God'', which we have not taken
into account here because they are expected to produce a negligible effect
at the scales we were using.

Furthermore, we have used some estimations of $y(z)$ given directly or indirectly by the literature, leading to the results in Table \ref{Tab:chi2} or in Figure \ref{Fig:models2}.
We observe that there are four cosmological models excluded within $>95$\% CL:
Einstein--de Sitter, open--Friedman cosmology without dark energy, flat quasi-steady-state cosmology (QSSC), and a static Universe with a linear Hubble law. Two models fit the data with a higher probability: concordance cosmology, and static Universe with tired light redshift.

For the standard concordance model ($\Omega =\Omega _m+\Omega _\Lambda =1$), we have set constraints on $\Omega _m$ to produce a better fit to the data; they are: $\Omega _m=0.24^{+0.10}_{-0.07}$, which is in agreement with values derived from other independent methods. Note, however, that, since we have not used any a priori assumptions, neither about the redshift distortion model nor other constraints including other data, the error bars here are larger than those derived from other methods---e.g., a recent analysis of CMBR data with {\it Planck} gives $\Omega _m=0.315\pm 0.017$ (Planck collaboration 2013). 
If we also allow a free value for the equation of state of the dark energy, $\omega _\Lambda $, instead of a fixed value of -1, we get as a best fit:
$\Omega _m=0.18^{+0.32}_{-0.16}$, $\omega _\Lambda=-1.2^{+0.7}_{-1.3}$.
Certainly, our method here is not the best one to
explore the possible values of the cosmological parameters once the cosmology is established; this could be much improved with further constraints, as Ross et al. (2007) and Okumura et al. (2008) did, but this was not our goal in this paper. We have focused on a method to test cosmological models without any a priori assumption.

The QSSC model is excluded at a $>95$\% CL, but the fit is better (excluded with a probability of 98.0\% instead of 99.56\%) with parameters $\Omega _m=1.22^{+0.12}_{-0.08}$ and  $\Omega _\Lambda =0^{+0}_{-0.08}$, which lead to $\Omega _c=1-\Omega _m-\Omega _\Lambda=-0.22^{+0.11}_{-0.12}$. This is compatible with the previously established value for the $C$-field density of $\Omega _c=-0.18$ (Banerjee \& Narlikar 1999, with the angular redshift test) or $\Omega _c=-0.27$ (Banerjee et al. 2000, with SNe data). The main problem with these fitted values is that they lead to the maximum redshift of an observable galaxy at $z\approx 4.5$, and we know there are galaxies beyond that. Anyway, a flat QSSC appears to have a very low probability to be a valid model according to only the present analysis.
All these considerations are for a flat QSSC universe; if we allowed curvature ($K\ne 0$), a wider range of probabilities could be obtained. 

The static models are not totally excluded from the present analysis. The one
with a linear Hubble law is excluded (assuming, as we did, that the linear regime applies and that the
peculiar velocity of the galaxies is irrotational, i.e., with zero vorticity, as predicted by the usual
gravitational growth theories), but not the one with tired-light hypothesis to explain the redshift. This tired-light model is also supported
by some other tests, such as the angular size test (L\'opez-Corredoira 2010;
LaViolette 2012, \S 7.4) or differential galaxy number counts versus magnitude (LaViolette 2012, \S 7.7).
Fitting the Hubble diagram as
$(1+z)=\exp\left(k\times \frac{D_L}{(1+z)^r}\right)$, with $D_L$ being the luminosity distance, is not good with SNe data using a ``simple'' tired light assumption (where $r=1/2$; the factor $(1+z)^{1/2}$ stems from the fact that the luminosity is proportional to $D_L^2$ and is inversely proportional to $(1+z)$ due to the redshift without expansion, i.e., without time dilation). However, the fit is acceptable using a ``plasma'' tired light redshift ($r=3/2$; L\'opez-Corredoira 2010), and it is quite good with gamma-ray bursts even up to $z=8$ with a tired-light model of type $r=1$ (Marosi 2013).

Future spectroscopic surveys with significantly more sources, coverage, and depth, will be able to improve the present results. In particular,
at high redshift we still have few spectroscopic QSOs, so the error bars are large (see how noisy the correlations are in Figs. \ref{Fig:elipses2Q}, \ref{Fig:elipses2Q2}). One exception is the point from Da \^Angela et al. (2005b) at $z=3.0$: $\Omega _m=0.35^{+0.65}_{-0.22}\ \longrightarrow \ y=2.26\pm 0.40$ (Note, however, that, looking at Figure 5 of Da \^Angela et al. (2005b), one could say that $\Omega _m>1$ is quite likely within 1$\sigma $, and consequently the error bars might be somewhat larger). As it can be observed in Figure \ref{Fig:models}, accurate values of $y$ at $z\gtrsim 2.5$ will be able to discern between the two successful cosmological models in this paper: $\Lambda $CDM (concordance) and static model with tired-light redshift. Also, very accurate measurements of $y$ at low $z$ would be able to test which one is the correct, since $\lim _{z\rightarrow 0}y(z)\approx 1+\frac{3}{4}\Omega _mz$ for the concordance model, whereas $\lim _{z\rightarrow 0}y(z)\approx 1+\frac{1}{2}z$
for the tired-light one. It is also worth noting that the most important
contribution stems from the already available data from previous analyses
of the anisotropic correlation function, combining the results of six references. The
contribution of the  data ($N=7$) analyzed in this paper does not change significantly the
confidence levels of exclusion of the different cosmological models because we get higher error bars. 
Note, however, that we cannot guarantee that the results of those six references with somewhat lower error bars are correct. In principle, from the experience developing this paper, the conclusion that comes up is that the quasi-degeneracy between the values of $\beta $ and $y$ for the fit of $\xi (\Delta z, z\Delta \theta ;z)$ is a
severe problem which produces huge error bars of $y$ and, consequently, the Alcock \& Paczy\'nski method only becomes powerful when we join the results from different surveys.

\

\acknowledgements  
{\bf Acknowledgements:} Thanks are given to: Juan E. Betancort-Rijo (IAC, Tenerife, Spain) for discussions on the methodology of the Alcock-Paczy\'nski test, Donald Schneider (SDSS-III Scientific Publications Coordinator) for his recommendations to bring the paper into compliance with the SDSS-III Publications Policy, and to the anonymous referee for helpful comments.

Funding for the SDSS-II has been provided by the Alfred P. Sloan Foundation, the Participating Institutions, the National Science Foundation, the U.S. Department of Energy, the National Aeronautics and Space Administration, the Japanese Monbukagakusho, the Max Planck Society, and the Higher Education Funding Council for England. The SDSS Web Site is http://www.sdss.org/.
The SDSS is managed by the Astrophysical Research Consortium for the Participating Institutions. The Participating Institutions are the American Museum of Natural History, Astrophysical Institute Potsdam, University of Basel, University of Cambridge, Case Western Reserve University, University of Chicago, Drexel University, Fermilab, the Institute for Advanced Study, the Japan Participation Group, Johns Hopkins University, the Joint Institute for Nuclear Astrophysics, the Kavli Institute for Particle Astrophysics and Cosmology, the Korean Scientist Group, the Chinese Academy of Sciences (LAMOST), Los Alamos National Laboratory, the Max-Planck-Institute for Astronomy (MPIA), the Max-Planck-Institute for Astrophysics (MPA), New Mexico State University, Ohio State University, University of Pittsburgh, University of Portsmouth, Princeton University, the United States Naval Observatory, and the University of Washington.

Funding for SDSS-III has been provided by the Alfred P. Sloan Foundation, the Participating Institutions, the National Science Foundation, and the U.S. Department of Energy Office of Science. The SDSS-III web site is http://www.sdss3.org/.
SDSS-III is managed by the Astrophysical Research Consortium for the Participating Institutions of the SDSS-III Collaboration including the University of Arizona, the Brazilian Participation Group, Brookhaven National Laboratory, University of Cambridge, Carnegie Mellon University, University of Florida, the French Participation Group, the German Participation Group, Harvard University, the Instituto de Astrof\'\i sica de Canarias, the Michigan State/Notre Dame/JINA Participation Group, Johns Hopkins University, Lawrence Berkeley National Laboratory, Max Planck Institute for Astrophysics, Max Planck Institute for Extraterrestrial Physics, New Mexico State University, New York University, Ohio State University, Pennsylvania State University, University of Portsmouth, Princeton University, the Spanish Participation Group, University of Tokyo, University of Utah, Vanderbilt University, University of Virginia, University of Washington, and Yale University. 

\

\appendix

\section{Values of the functions $x(z)$ and $y(z)$ for different cosmological models}
\label{.cosmomodels}

The comoving distance between two sources separated by a relatively small
$\Delta z$ and $\Delta \theta $ (radians) in the redshift space is
\begin{equation}
s=\sqrt{\left[\frac{d[d_{\rm com}(z)]}{dz}\ \Delta z\right]^2\ +\ [(1+z)^md_{\rm A}(z)\ \Delta \theta ]^2}
\end{equation}\[
=\frac{c}{H_0}x(z)\sqrt{(\Delta z)^2\ +\ y^2(z)\ (z\Delta \theta )^2}
,\]
where $d_{\rm A}$ and $d_{\rm com}$ are our angular and comoving distances, respectively, from the first source, $m=1$ with
expansion or 0 without expansion, and
\begin{equation}
x(z)\equiv \frac{H_0}{c}\frac{d[d_{\rm com}(z)]}{dz}
,\end{equation}
\begin{equation}
y(z)\equiv \frac{H_0}{c}\frac{(1+z)^md_{\rm A}(z)}{z\ x(z)}
,\end{equation}
with $H_0$ as the Hubble constant, and $c$ as the speed of light.

Here we give these dependences for some cosmological models.

\begin{enumerate}

\item Standard concordance model: $m=1$,
\begin{equation}
\label{concordance}
d_{\rm com}(z)=(1+z)d_{\rm A}(z)=
\end{equation}
\[
\frac{c}{H_0}
\int _0^z\frac{dx}{\sqrt{\Omega _m(1+x)^3
+\Omega _\Lambda (1+x)^{3(1+\omega _\Lambda )}}}
,\] 
where $\Omega _m$ and $\Omega _\Lambda $ are, respectively, the density of matter and dark energy
and $\omega _\Lambda $ is the dimensionless number associated with the equation of state of the
dark energy. We assume that these parameters do not vary with $z$ and that the universe is flat, i.e., 
$\Omega _m+\Omega _\Lambda =1$. Hence,

\begin{equation}
x(z)=\frac{1}{\sqrt{\Omega _m(1+z)^3+\Omega _\Lambda (1+z)^{3(1+\omega _\Lambda )}}}
,\end{equation}
\begin{equation}
y(z)=\frac{1}{z}\int _0^zdx \sqrt{ \frac{\Omega _m(1+z)^3
+\Omega _\Lambda (1+z)^{3(1+\omega _\Lambda )}} {\Omega _m(1+x)^3
+\Omega _\Lambda (1+x)^{3(1+\omega _\Lambda )} } }
.\end{equation}
For very low $z$,
\begin{equation}
\lim _{z\rightarrow 0}y(z)\approx 1+
\frac{1}{4}[3\Omega _m+3(1+\omega _\Lambda )\Omega _\Lambda ]z
\label{lowz_1}
.\end{equation}

For the standard model, we adopt the values of $\Omega _m=0.3$, $\Omega _\Lambda =1-\Omega _m=0.7$, and
$\omega _\Lambda =-1$.

\item Einstein--de Sitter model (Equation (\ref{concordance}) with $\Omega
_\Lambda =0$, $\Omega _m=1$, and $\omega _{\Lambda }=-1$): $m=1$,
\begin{equation}
d_{\rm com}(z)=(1+z)d_{\rm A}(z)=\frac{2c}{H_0}\left[1-\frac{1}{\sqrt{1+z}}\right]
,\end{equation}
\begin{equation}
x(z)=\frac{1}{(1+z)^{3/2}}
,\end{equation}
\begin{equation}
y(z)=\frac{2}{z}\left[(1+z)^{3/2}-(1+z)\right]
.\end{equation}
For very low $z$,
\begin{equation}
\lim _{z\rightarrow 0}y(z)\approx 1+\frac{3}{4}z
.\end{equation}

Although this is not the standard model nowadays, there are
some researchers who still consider it more appropriate than the concordance
model (e.g., Vauclair et al. 2003; Blanchard 2006; Andrews 2006).
Regarding the compatibility with Type Ia SNe data, see discussion in L\'opez-Corredoira
(2010, \S 5.3).

\item Friedmann model of negative curvature with $\Omega _m=0.3$ and  
$\Omega _\Lambda=0$, which implies a curvature term $\Omega _K=1-\Omega _m=0.7$
(Baryshev \& Teerikorpi 2012, \S 7.4.1): $m=1$,
\begin{equation}
d_{\rm com}(z)=\frac{c}{H_0}
\int _0^z\frac{dx}{\sqrt{\Omega _m(1+x)^3
+\Omega _K(1+x)^2 }}
,\end{equation}
\begin{equation}
d_{\rm A}(z)=\frac{c}{(1+z)H_0\sqrt{\Omega _K}}
\end{equation}\[
\times \sinh \left(\sqrt{\Omega _K}\int _0^z\frac{dx}{\sqrt{\Omega _m(1+x)^3
+\Omega _K(1+x)^2 }}\right)
,\]
\begin{equation}
x(z)=\frac{1}{\sqrt{\Omega _m(1+z)^3+\Omega _K(1+z)^2}}
,\end{equation}
\begin{equation}
y(z)=\frac{\sqrt{\Omega _m(1+z)^3+\Omega _K(1+z)^2}}{z\sqrt{\Omega _K}}
\end{equation}\[
\times \sinh\left[\sqrt{\Omega _K}\int _0^z\frac{dx}{\sqrt{\Omega _m(1+x)^3
+\Omega _K(1+x)^2}}\right]
.\]
For very low $z$,
\begin{equation}
\lim _{z\rightarrow 0}y(z)\approx 1+\frac{1}{4}(3\Omega _m+2\Omega _K)z
.\end{equation}

This model might be considered for the case where we accept that $\Omega _m\approx 0.3$, but we
do not want to include any $\Lambda $ term.

\item Quasi-steady state cosmology (QSSC), 
$\Omega _m=1.27$, $0\ge \Omega _\Lambda =-0.09$, and $\Omega _c=1-\Omega _m-\Omega _\Lambda =-0.18$ 
($C$-field density; Banerjee \& Narlikar 1999): $m=1$,
\begin{equation}
d_{\rm com}(z)=(1+z)d_{\rm A}(z)=\frac{c}{H_0(1+z)}
\end{equation}\[
\times \int _0^z\frac{dx}{\sqrt{\Omega _c(1+x)^4+\Omega _m(1+x)^3
+\Omega _\Lambda }}
,\]
\begin{equation}
x(z)=\frac{1}{\sqrt{\Omega _c(1+z)^4+\Omega _m(1+z)^3+\Omega _\Lambda }}
,\end{equation}
\begin{equation}
y(z)=\frac{1}{z}\int _0^zdx \sqrt{\frac{\Omega _c(1+z)^4+\Omega _m(1+z)^3
+\Omega _\Lambda }{\Omega _c(1+x)^4+\Omega _m(1+x)^3
+\Omega _\Lambda }}
.\end{equation}
For very low $z$,
\begin{equation}
\lim _{z\rightarrow 0}y(z)\approx 1+\frac{1}{4}(4\Omega _c+3\Omega _m)z
.\end{equation}

This cosmology is not the standard model, but it can also fit
many data on some cosmological tests (Banerjee \& Narlikar 1999;
Banerjee et al. 2000; Narlikar et al. 2002; Vishwakarma 2002).
The expansion with an oscillatory term gives a dependence
of the luminosity and angular distance similar to the standard model,
adding the effect of matter creation ($C$-field) with 
slight changes depending on the parameters.
The parameters of this cosmology are not as well constrained as
those in the standard model. Here, I use the
best fit for a flat ($K=0$) cosmology given by Banerjee \& Narlikar (1999):
$\Omega _m=1.27$, $\Omega _\Lambda =-0.09$, and $\Omega _c=-0.18$, which
corresponds to $\eta =0.887$ (amplitude of the oscillation relative to 1), 
$x_0=0.797$ (ratio between the actual size of the universe and the average
size in the present oscillation), and the maximum allowed redshift of a galaxy
$z_{\rm max}=6.05$ (Note, however, that the maximum observed redshift has 
risen above 8 nowadays according to some authors, e.g., Lehnert et al. 2010). Other 
preferred sets of parameters give results that are close.
The  values most used are $K=0$, $\Omega _\Lambda =-0.36$, $\eta =0.811$,
and $z_{\rm max}=5$ (Banerjee et al. 2000; Narlikar et al. 2002; Vishwakarma 2002; Narlikar et al. 2007), 
which imply $\Omega _m=1.63$ and $\Omega _c=-0.27$,
but I avoid them because they do not allow  galaxies to be fitted with $z>5$.
Parameters with a curvature different from zero ($K\ne 0$) also give results
that are very close in the angular size test (Banerjee \& Narlikar 1999).

\item Static Euclidean model with linear Hubble law for all redshifts: $m=0$,
\begin{equation}
d_{\rm com}(z)=d_{\rm A}(z)=\frac{c}{H_0}z
,\end{equation}
\begin{equation}
x(z)=1
,\end{equation}
\begin{equation}
y(z)=1
.\end{equation}

These simple relations 
indicate that  redshift is always proportional
to  angular distance, i.e., a Hubble law. We assume in this scenario that the
universe is static.
The caveat is to explain
the mechanism as different from the expansion or the Doppler effect, which gives rise
to the redshift. This cosmological model is not a solution that
has been explored theoretically or mathematically.
However, from a phenomenological point of view, we can 
consider this relationship between distance and redshift
as an ad hoc extrapolation from the observed dependence on the low-redshift
universe.

\item Tired-light static Euclidean model: $m=0$,
\begin{equation}
d_{\rm com}(z)=d_{\rm A}(z)=\frac{c}{H_0}{\rm ln} (1+z)
\label{angdisttl}
,\end{equation}

\begin{equation}
x(z)=\frac{1}{(1+z)}
,\end{equation}
\begin{equation}
y(z)=\ln (1+z)\left[1+\frac{1}{z}\right]
.\end{equation}
For very low $z$,
\begin{equation}
\lim _{z\rightarrow 0}y(z)\approx 1+\frac{1}{2}z
.\end{equation}

This is again a possible ad hoc phenomenological representation
which stems from considering that the photons
lose energy along their paths due to some interaction, and the
relative loss of energy is proportional to the length of that path
(e.g. LaViolette 1986; Brynjolfsson 2004a, Section 5.8; Ashmore 2011;
LaViolette 2012, Section 7.3), i.e., $\frac{dE}{dr}=-\frac{H_0}{c}E$. 

\end{enumerate}

\section{Common pairs in two adjacent subsamples for the calculation of the
Two-point correlation function}
\label{.2subsamples_tpcf}

\begin{figure*}[htb]
\vspace{1.2cm}
{\par\centering \resizebox*{16cm}{10.5cm}{\includegraphics{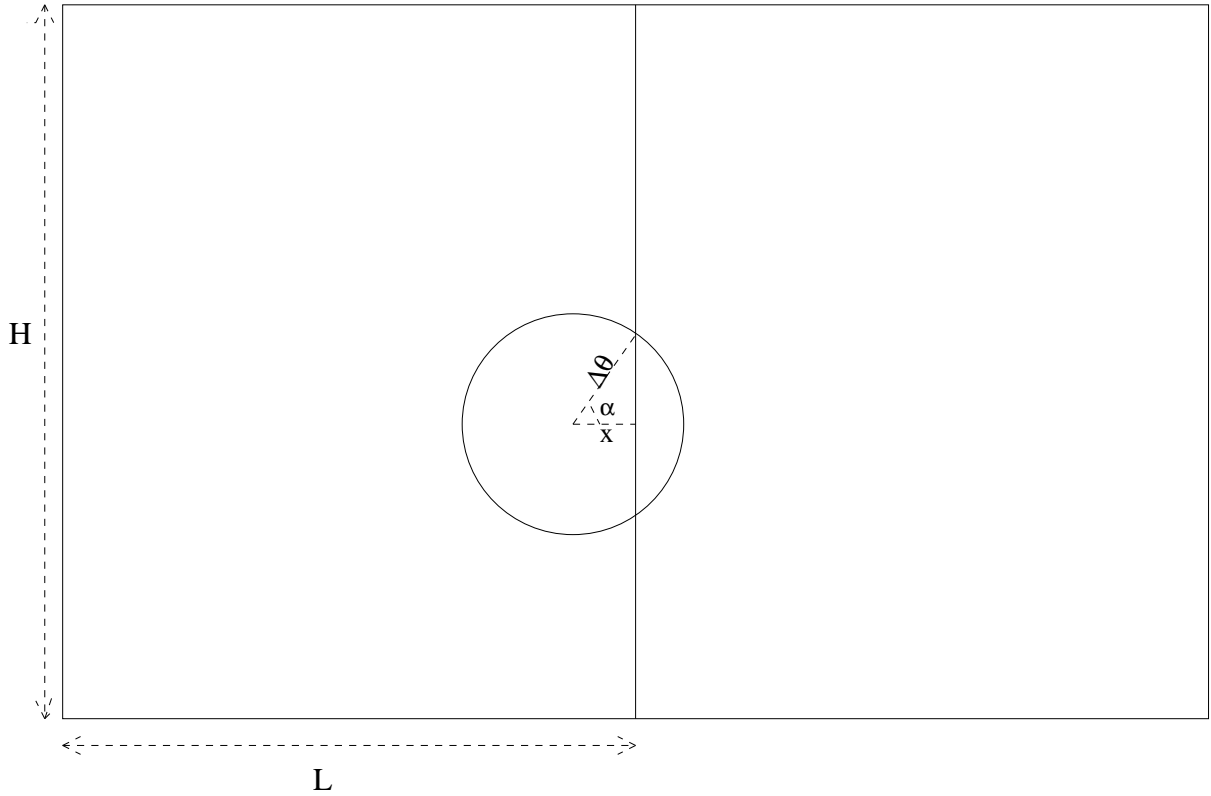}}
\par\centering}
\caption{Graphical representation of two adjacent rectangular regions for which we want to calculate
the probability of having pairs of objects separated by an angular distance $\Delta \theta $ 
with one of them in each region.}
\label{Fig:2subsamples_tpcf}
\end{figure*}

Given two rectangular adjacent regions where we want to calculate the two-point correlation function of parent objects embedded in each region within an angular distance $\Delta \theta $, we want to calculate the number of common pairs in both subsamples. Both regions have a width, $L$, and height, $H$, and they are in contact through the vertical divisory line, as shown in Figure \ref{Fig:2subsamples_tpcf}. Given any parent source in the left box at any height with completely covered circle, the probability that the second object of the pair within an angular distance $\Delta \theta $ is in the right box is:
\begin{equation}
P=\frac{1}{L}\int _0^{\Delta \theta }dx\ f(x)
,\end{equation}
where $f(x)$ is the fraction of the circle in the right box, according to Figure \ref{Fig:2subsamples_tpcf}:
\begin{equation}
f(x)=\frac{\alpha }{\pi }
,\end{equation}\[
\cos \alpha =\frac{x}{\Delta \theta }
.\]
Hence,
\begin{equation}
P=\frac{1}{\pi L}\int _0^{\Delta \theta }dx\ \cos ^{-1}\left(\frac{x}{\Delta \theta }\right)=\frac{\Delta \theta }{\pi L}
.\end{equation}

Therefore, the number of common pairs separated by an angular distance $\Delta \theta $ is $\frac{\Delta \theta }{\pi L}$. This gives us the degree of dependence of the calculation of the two-point correlation function in both adjacent subsamples.

If we took the horizontal direction as the right ascension and the vertical direction as the declination (as will be done in this paper), and the whole sky is divided into $N_f$ subsamples, each pair of adjacent subsamples would have an average $\Delta R.A=\frac{2\pi }{N_f}$. The subregions would not be rectangular but slices of constant R.A. width; this
can be translated into an average $\langle L\rangle =(\Delta {\rm R.A.})\frac{\pi }{4}$ ($\pi /4$ stems from
$\frac{\int _0^{\pi/2}d\delta\,\cos^2\delta }{\int _0^{\pi/2}d\delta\,\cos \delta }$). Thus, the average number of
common pairs in adjacent regions would be
\begin{equation}
P\sim \frac{2N_f\Delta \theta }{\pi ^3}
.\end{equation} 
For $\Delta \theta <L$, there are only common pairs among the adjacent subsamples, not with the regions that are not in contact; therefore if we select two random subsamples, the probability of a pair to have a common pair in both subsamples is $P$ multiplied by a factor $\frac{2}{N_f-1}$ (2 is the number of adjacent regions, and $N_f-1$ are the total number of subsamples except the first selected one). We neglect the pairs near the polar caps, which are very few in comparison with the rest of the pairs; in the application of the present paper, there
are no SDSS galaxies near the polar cap so their contribution is almost null. 
Therefore, the degree of correlation among subsamples (which would generate some terms in the covariance matrix) is roughly (we approximate $N_f-1\approx N_f$ for high values of $N_f$)
\begin{equation}
P\sim \frac{4}{\pi^3}\Delta \theta
,\end{equation}
an amount which is much lower than the one for low $\Delta \theta $; it can be 
considered negligible ($P\lesssim 0.05$) for $\Delta \theta \lesssim 20^\circ $.

\section{Two-point correlation function of a random distribution with mean density
variable with redshift}
\label{.varrho}

From the definition, a Poisson distribution of sources with constant mean density in the space gives a null two-point correlation function. But if there is some variation of the mean density with redshift, even locally preserving the random/Poisson distribution of sources, it will give a non-null contribution:

\begin{equation}
\xi ^{0}(\Delta z, z\Delta \theta ;z)=\frac{1}{2}\sum _{m=-1,1}\left[\frac{\langle \rho (z)\rho (z+m\Delta z)\rangle }{\langle \rho (z)\rangle ^2}-1\right]
.\end{equation}
Using a Taylor series,
\begin{equation}
\rho (z+m\Delta z)=\rho (z)
\end{equation}\[
\times \left[1+m\frac{\rho '(z)}{\rho (z)}(\Delta z)
+\frac{m^2}{2}\frac{\rho ''(z)}{\rho (z)}(\Delta z)^2+O[(m\Delta z)^3]\right]
,\]
we get
\begin{equation}
\xi ^{0}(\Delta z, z\Delta \theta ;z)=\frac{1}{2}\frac{\rho ''(z)}{\rho (z)}(\Delta z)^2+O[(\Delta z)^4]
\end{equation}

For the isotropic case,
\begin{equation}
\xi ^0(r;z)=\frac{1}{2}\int _0^\pi d\theta \sin \theta
\left[\frac{\langle \rho (z)\rho (z+r\cos \theta)\rangle }{\langle \rho (z)\rangle ^2}-1\right]
.\end{equation}
Substituting $\rho (z+r\cos \theta)$ with a Taylor series like above,
\begin{equation}
\xi ^0(r;z)=\frac{1}{6}\frac{\rho ''(z)}{\rho (z)}r^2+O[r^4]
.\end{equation}

\end{document}